\documentclass[a4paper,twoside,10pt]{letter}
\usepackage{graphicx,saj,multicol,subeqnarray}

\def\papertype{Preliminary report}

\begin{document}
\baselineskip=3.1truemm
\columnsep=.5truecm
\newenvironment{lefteqnarray}{\arraycolsep=0pt\begin{eqnarray}}
{\end{eqnarray}\protect\aftergroup\ignorespaces}
\newenvironment{lefteqnarray*}{\arraycolsep=0pt\begin{eqnarray*}}
{\end{eqnarray*}\protect\aftergroup\ignorespaces}
\newenvironment{leftsubeqnarray}{\arraycolsep=0pt\begin{subeqnarray}}
{\end{subeqnarray}\protect\aftergroup\ignorespaces}
%

% Running titles

\markboth{\eightrm HABITABILITY OF GALAXIES AND THE APPLICATION OF MERGER TREES IN ASTROBIOLOGY} {\eightrm N. STOJKOVI{\' C}, B. VUKOTI{\' C} and M.M. {\' C}IRKOVI{\' C}}

{\ }

%\publ

%\type

{\ }

% Title

\title{HABITABILITY OF GALAXIES AND THE APPLICATION OF MERGER TREES IN ASTROBIOLOGY}

% Authors

\authors{N. Stojkovi{\' c}$^{1}$, B. Vukoti{\' c}$^{2}$ and M.M. {\' C}irkovi{\' c}$^2$}

\vskip3mm

% Address

\address{$^1$Department of Astronomy, Faculty of Mathematics,
University of Belgrade\break Studentski trg 16, 11000 Belgrade,
Serbia}

\Email{nstojkovic}{aob.rs}

\address{$^2$Astronomical Observatory Belgrade, Volgina 7,  11060 Belgrade, Serbia}

\Email{bvukotic}{aob.rs, mcirkovic@aob.rs {\normalfont \small(corresponding author)}}

% Received and Accepted dates

%\dates{????}{????}

% Abstract

\summary{Galaxies represent the main form of organization of matter in our universe. Therefore, they are of obvious interest for the new multidisciplinary field of astrobiology. In particular, to study habitability of galaxies represents one of the main emerging challenges of theoretical and numerical astrobiology. Its theoretical underpinnings are, however, often confused and vague. Here we present a systematic attempt to list and categorize major causal factors playing a role in emergent habitability of galaxies. Furthermore, we argue that the methodology of cosmological merger trees is particularly useful in delineating what are systematic and lawful astrobiological properties of galaxies at present epoch vs. those which are product of historical contingency and, in particular, interaction with wider extragalactic environment. Employing merger trees extracted from cosmological N-body simulations as a new and promising research method for astrobiology has been pioneered by Stanway et al. (2018). We analyse the general issue of applicability of merger trees and present preliminary results on a set of trees extracted from the Illustris Project. In a sense, this approach is directly complementary to using large-scale cosmological simulations to study habitable zones of individual galaxies with high mass/spatial resolution; taken together they usher a new era of synergy and synthesis between cosmology and astrobiology. }

% Keywords (see keywords.pdf file)

\keywords{galaxies: evolution -- astrobiology -- methods: numerical -- fine tuning -- galaxies: evolution -- extraterrestrial intelligence}

%\begin{multicols}{1}
{

% Sections

\section{1. INTRODUCTION: COSMOLOGY AND HABITABILITY}

The concept of habitability has gradually become the key concept of the revolutionary multidisciplinary field of astrobiology (for general reviews see e.g., Des Marais and Walter 1999; Chyba and Hand 2005; Horneck and Rettberg 2007; Des Marais et al. 2008; Hanslmeier 2009; Seager 2013; Domagal-Goldman et al. 2016; Cockell et al. 2016). Since astrobiology is such a rapidly advancing field, it is impossible to have a clear, formal definition of its concept, the situation often encountered in immature fields of science (see, for instance, examples from the early history of cosmology discussed in {\' C}irkovi{\' c} 2012). In contrast, as with the definition of life itself, there are multiple operational definitions used often implicitly by particular researchers and schools of thought; we shall consider some examples of this further on. Also worthy of notice are attempts to create various scales or scorecards of numerical values for habitability (e.g., Franck et al. 2001; Shock and Holland 2007; Gowanlock, Patton, and McConnell 2011; Barnes, Meadows, and Evans 2015; Shields, Ballard, and Johnson 2016; Bora et al. 2016; Rodríguez-Mozos and Moya 2017; Jagadeesh et al. 2017, 2018; Saha et al. 2018) of practical value in characterizing large numbers of extrasolar planets. 

In light of the enormous advances achieved in discovering and studying properties of other planetary systems, it is entirely natural that the notion of habitability is the most relevant as defined for planets, in particular Earth. Almost all studies cited pertain to attempts to characterize habitability of planets, together with the recently studied possibilities of habitable moons (Heller 2012; Heller et al. 2014). The notion of the circumstellar habitable zone (CHZ) codified the standard, geocentric, and anthropocentric view of planetary habitability. It has become problematic within the Solar System already, with the insights which led researchers to consider habitability of subglacial water oceans like those on Europa and Enceladus, or even ``prebiotic chemistry analogues'' like shallow methane ponds on the surface of Titan (e.g., Lunine 2009; McKay et al. 2014; Mitri et al. 2018). Those outer Solar System objects definitely do not belong to Sun's CHZ and yet, they have become completely legitimate targets of astrobiological investigations. However, even if habitability were a perfectly well-defined notion within a planetary system, this would have still prompted two important questions:
\vspace{0.5cm}
\item{1.} What external influences (i.e., outside of the given planetary system) are capable of significantly impacting planetary habitability?
\item{2.} Can we meaningfully discuss habitability at higher levels of organization of matter (e.g., at levels of stellar clusters, stellar populations, galaxies and various galaxy groups and clusters)?
\vspace{0.5cm}

\noindent The questions are intertwined, since elementary astronomy clearly tells us that planetary systems are not isolated, ``closed box'' systems whose formation and evolution could be studied in isolation from their wider environment. On the contrary, there is a wealth of evidence that both the number and nature of planetary systems are strongly correlated with both their spatiotemporal position and astrophysical and astrochemical properties of their environment. Conversely, once we understand specifics of the wider environment of the star we may, at least in vague outline, get some insight into habitability of its planetary or satellite companions.

The idea is not entirely new. The first person to argue that habitability is (at least to a degree) influenced by the cosmic environment outside of our own Solar System was the co-discoverer of natural selection and a key precursor of modern astrobiology Alfred Russel Wallace, who in 1903 published what is arguably the first astrobiological monograph, Man's Place in the Universe. Among other thing which were truly and outstandingly original at the time, Wallace argued that, in agreement with the best observational astronomy of his time, Sun and the Solar System are located near the centre of our stellar system, the Milky Way, in a cluster of stars somewhat offset from the exact centre, but conductive to origin and evolution of life. He wildly speculated about the exact mechanism of such an influence of the ``central cluster'' on habitability, including a somewhat romantic idea that starlight exerts beneficial effects on abiogenesis (Wallace 1903). In spite of the fact that his empirical predictions were spectacularly falsified after discoveries of Shapley, Hubble, and others about our position in the Milky Way, and Milky Way's position as just one of billions of galaxies in the observable universe, Wallace was definitely the first to ask the right questions and to argue for a kind of galactic habitability: the distribution of habitable sites within our stellar system is highly non-uniform. Using the wrong cosmological model prevented Wallace from reaching realistic conclusions, but his questions were right on target and provided a model for all subsequent inquiries into cosmological aspects of habitability.

The modern epoch of studying of galactic habitability has been inaugurated with the appearance, in 2000, of the ``rare Earth hypothesis'' of Peter Ward and Donald Brownlee, arguably the first general astrobiological theoretical framework (Ward and Brownlee 2000). In a nutshell, the rare-Earth hypothesis is a probabilistic argument suggesting that, while simple microbial life is probably existent throughout the Galaxy, complex biospheres like ours are very rare due to the exceptional combination of many distinct requirements. These ingredients of the rare-Earth hypothesis are well known to even a casual student of astrobiology:
\vspace{0.5cm}
\item{$\bullet$} \textsl{Galactic Habitable Zone}: A habitable planetary system needs to be in a narrow annular ring in the Milky Way disc, where chemical abundances and stability conditions for the formation of habitable planets, as well as the emergence and evolution of life are satisfied.
\item{$\bullet$} \textsl{Circumstellar habitable zone}: A habitable planet needs to be in the narrow interval of distances from the parent star in order to possess surface water.
\item{$\bullet$} \textsl{Rare Moon}: Having a large moon to stabilize the planetary axis is crucial for long-term climate stability of a habitable planet.
\item{$\bullet$} \textsl{Rare Jupiter}: Having a giant planet (`Jupiter') at the right distance to deflect much of incoming cometary and asteroidal material enables a sufficiently low level of impact catastrophes.
\item{$\bullet$} \textsl{Rare nuclides}: Radioactive r-elements need to be present in the planetary interior in sufficient amounts to enable plate tectonics and the functioning of the carbon--silicate cycle.
\item{$\bullet$} \textsl{Rare Cambrian-explosion analogues}: The evolution of complex metazoans requires exceptional physical, chemical and geological conditions for episodes of sudden diversification and expansion of life.
\vspace{0.5cm}

\noindent While there is considerable debate about the significance of these and other individual items, the key innovation is obviously given by the first one: the concept of the Galactic Habitable Zone (henceforth GHZ) which Ward and Brownlee introduced together with Guillermo Gonzalez (Gonzalez, Brownlee, and Ward 2001). It is simply the region of the Galaxy in which we expect to find habitable planets in the sense of possessing liquid water on the surface and provide stable, long-term environment for metazoan life. In subsequent construals, it has been considered as a region where the probability of habitable planets is non-negligible. This already points to a possible weakening of the original concept, which is now assigned a probabilistic interpretation. Of course, a number of assumptions were built in the original formulation of GHZ which could be validated only in the course of detailed numerical and observational work. In particular, Gonzalez et al. (2001) argued that metallicity is the major controlling parameter determining the width of GHZ which was conceived as an annular ring within the Milky Way thin disk. In the innermost regions and within the Galactic bulge, habitability is limited by processes such as dynamical instability of planetary systems and the frequency of supernova irradiations, but once we leave this central region, the width of the ring is determined solely by the metallicity gradient in the disk; habitable zone extends as far as metallicity is greater than the critical value considered necessary for the formation of terrestrial planets. This formulation left a minority ($\sim10\%$) of the stellar population of the Milky Way inside GHZ -- still a huge number, but in general agreement with the rare Earth anti-Copernican bias.

This original formulation has obviously left room for generalization to external galaxies as well, since it referenced processes and trends like the supernova irradiations or metallicity gradients which have been observed as general features of galactic evolution and not some peculiar features of the Milky Way. However, in the original rare-Earth framework, habitability of external galaxies has been given a rather simple form: it was argued by Ward and Brownlee (2000, especially pp. 29-31) that only giant, metal-rich spirals like the Milky Way possess sizeable GHZs, and therefore could harbour life. It was also understood that even within this category of galaxies, the relevant extragalactic GHZs are all annular rings determined by local metallicity gradients. 

This is one of the rare-Earth conjectures which has proved spurious, to say the least, since the turn of the century. In particular, it is clear now that the early samples of extrasolar planets discovered mostly by spectroscopic technique have been heavily biased toward larger metallicities. In other words, the importance of metallicity as the discriminator between what is habitable and what is not has declined when more representative samples have become available in the Kepler-era. We shall return to this topic below. Before that, it is important to notice that the interest in habitability of external galaxies has been reawakened recently by several interrelated developments which undermined the rare-Earth paradigm. For instance, the work of Suthar and McKay (2012) indicated that the low-metallicity wing of the extrasolar planet metallicity distribution is perfectly in accordance with the possibility that such planets exist in elliptical galaxies such as M87 and M32.

In addition to testing of the rare-Earth claims, there are several other important motivations for studying the habitability of galaxies. Great successes of the current $\lambda$CDM cosmological models of structure formation (e.g., Weinberg 2008) have prompted a strong interest in applying large-scale cosmological numerical simulations to obtain better understanding in other astronomical fields. In addition, by being able to obtain both averages over ensemble and over time by any given epoch, we have solid grounds for being able to discriminate between what has been dynamical, systematic effects on habitability, vs. what is accidental, contingent, and historical happenstance in each particular case. The pioneering steps in applying cosmological simulations in astrobiology have been made by Vukoti{\' c} et al. (2016) and Forgan et al. (2017); for a review see also Vukoti{\' c} (2017). These studies shed some light on how the Galactic Habitable Zone emerges as a consequence of the large-scale trends of galactic evolution in isolated, or nearly-isolated giant galaxies essentially similar to the Milky Way. Both primary studies have shown a certain shift from the original range of galactocentric distances in Gonzalez et al. (2001) toward more outward regions of the disk, and hence \textsl{smaller mean metallicities}. For instance, in the last snapshot of Vukoti{\' c} et al. (2016), corresponding to the most recent epoch, the maximal probability of finding a habitable planet seems to be at galactocentric distances between 10 and 15 kpc. In addition, the study of Forgan et al. (2017) shows that dwarf galaxies, such as the satellites of Milky Way and Andromeda, can have a significant density of habitable systems, to even greater degree than their larger gravitational hosts. Both these effects frame a wider question to what extent galactic environments significantly diverging from the one in which Solar System has formed and evolved are conducive to life. Hence, these studies have indirectly supported the conclusion of Suthar and McKay (2012) that early-type systems could be habitable as well, in sharp contrast to the rare-Earth dogma. 

In part, this is a consequence of the dramatic shift of opinion about the role of metallicity of parent stars in formation and frequency of habitable planets (and planetary satellites) which has occurred recently. In contrast to early days of the extrasolar planets' discoveries -- which have been the days in which Gonzalez, Ward, and Brownlee has come up with the rare-Earth hypothesis -- ``hot Jupiters'' have been the most prominent feature discovered, and consistently high metallicity of their parent stars mislead many astrophysicists in concluding that this is a universal characteristic (e.g., Fischer and Valenti 2005, and references therein). When smaller planets began to be discovered, correlations with stellar metallicity began to wane (e.g., Buchhave et al. 2012), although some of it has remained, especially for M-dwarf parent stars (Schlaufman and Laughlin 2011). If the shift is real when we discuss stellar and planetary populations within a galaxy such as the Milky Way, it is only reasonable to generalize it to any consideration of any sample of galaxies as well: lower (average) metallicity galaxies will be more appreciated as potential habitats. 

This intuition has been confirmed by three recent external-galaxies-specific studies which have attracted lots of attention in this respect. Dayal et al. (2015) use a simple model which links global properties of galaxies with the factors limiting their habitability. In particular, the destructive effects of cosmic explosions (supernovae, gamma-ray bursts, and perhaps magnetars, extreme stellar flares, and similar events) have been the focus of the attention. It is clear that these effects can cause extreme ecological damage to local planetary biospheres and consequently reduce habitability within large volumes of space. For a sampling of the huge literature on the subject in the last more than 60 years, see Krasovsky and Shklovsky (1957); Dar, Laor, and Shaviv (1998); Scalo and Wheeler (2002); Gehrels et al. (2003); Thomas et al. (2005); Thomas and Melott (2006); Thomas (2009); Beech (2011); Marinho, Paulucci, and Galante (2014). Most cosmic explosions are proportional to the star formation rate, which gives us a handle to compare irradiated volumes and consequent decrease of habitability in different galaxies. This study separately tracks terrestrial planets, and gas-giants whose satellites might be habitable, as additive constituents of the overall galactic habitability. By comparing star-formation rates for a sample of SDSS galaxies, the authors conclude that depressing star-formation rate (and hence the irradiation rate) at recent epochs is the major factor of creating a large habitable volume. On that basis, they conclude that giant ellipticals with very small present-day star formation rates and moderately low metallicities, like Maffei 1, are the best abodes of life in the universe in the present epoch. Clearly, this is in sharp contrast to Ward and Brownlee (2000), Gonzalez et al. (2001), and other rare-Earth theorists.

The work of Gobat and Hong (2016) offers somewhat more cautious conclusions, on the basis of a semi-analytic model of galactic habitability which includes metallicity dependence and irradiations, as well as the finite duration of habitability of any planet due to the host star leaving the Main Sequence. They obtain a scaling of habitable fraction of planetary systems with the stellar mass of galaxies for various redshifts (epochs), as well as a rough dependence on the star formation rate. Their model indicates relatively minor effect of supernovae explosions, unless unrealistically large ``lethal radii'' are considered. On the other hand, Gobat and Hong emphasize the sensitivity of results on the stellar IMF, which is quite poorly constrained in the external galaxies. All in all, they conclude, in agreement with Dayal et al. (2015), that early-type (``passive'' in their terms) galaxies are, on the balance, more advantageous for the emergence and survival of life than star-forming (``active'') spirals like the Milky Way. 

Finally, Stanway et al. (2018) investigate the cosmic evolution of habitability with galaxy merger trees, the method naturally following the deployment of N-body simulations for astrobiological purposes by Vukoti{\' c} et al. (2016) and Forgan et al. (2017), which we shall discuss below in more detail. Stanway et al. use the Millenium Simulation and its derivative ``millimil'' sample in 63 time intervals (or ``snaps'') starting from the earliest redshifts to the present day. In addition to metallicity weighting, irradiations of potentially habitable planets by SNe, GRBs and the central AGN are considered in order to reach volume-averaged habitability over prolonged intervals of time (like the fiducial value of 6 Gyr). 

Metallicity plays a role in the model of Stanway et al. in two ways. First, there is a simple cut-off related to the terrestrial planet formation: regions with $Z/Z_\odot$ smaller than a threshold value are dropped from the total tally of habitable volume. This might turn out to be problematic, since it strongly depends on the physics adopted in the underlying simulation, which is of necessity much cruder in determining metallicities, and especially their local fluctuations, than the dedicated models of chemical evolution. Second, the same study correctly emphasizes that strong metallicity dependence of the core-collapse SNe/GRB rates presents an important, and possibly dominant, detrimental factor from the point of view of habitability, especially for the most massive (= the highest metallicity) systems.

All in all, Stanway et al. (2018) find a complex situation, with no simple answer to the question which galaxies are most habitable at present. Even galaxies of the same mass manifest clear differences regarding their possible habitability histories (changes in the habitable fraction of stellar mass over time). This study was the first to indicate that the distribution of habitable fraction among stellar systems is bimodal, being either smaller than $5\%$ in some galaxies, or greater than about $30\%$ in the other systems, when evaluated at present epoch. One way in which conclusions of Stanway et al. could be generalized upon is not relying upon the strong mass-metallicity relation imposed on objects in the Millenium Simulation, esp. in the low mass range (more on this below). 

These pioneering studies have been covered here in some detail not only because they are relevant for the rest of this manuscript, but also because they highlight some of the problems and pitfalls confronting any research in habitability of galaxies. One of the reasons why the topic is so complex and confusing lies -- of necessity -- in the domain of epistemology and philosophy of science. Habitability of galaxies has the hallmark traits of an \textsl{emergent phenomenon}. Rapid development of galactic astronomy has led us to the picture of galaxies as very complex systems in which properties of the whole cannot be fully reduced to properties of the individual components. It is not just accidental that researchers have been considering analogies and metaphors from terrestrial systems in speaking about, for example, recycling of interstellar and intergalactic medium (Duc, Braine, and Brinks 2004), ``galaxy ecology'' (Balogh et al. 2004), or ``rain'' (Heitsch and Putman 2009). Even these, purely astrophysical, processes are nowadays modelled with the degree of detail which makes intuitive and analytic approaches almost unfeasible. There are tantalizing observational indications that most or all galaxies could be arranged along a universal baryonic mass function (e.g., Jovanovi{\' c} 2017), which would introduce some underlying order into this mess of complex nonlinear interactions which shape phenomena on the galactic scale. When we move from \textsl{astrophysics} of galaxies to \textsl{astrobiology} of galaxies, things become even more involved, since the relevant feedbacks are not entirely understood in the terrestrial context either. Clearly, the complexity of the topic of habitability of galaxies is such that at least some rudimentary taxonomic work is required in order to make sense of each new result available. 

Therefore, the goals of the present paper are (i) to give a detailed and up-to-date overview of the controversial topic of habitability of galaxies, as the main ingredients of the structure of the universe at large scales; (ii) to present and justify the method of merger trees in studying galactic habitability as a quite new and potentially extraordinarily fruitful approach to this complex and involved problem; and (iii) to present preliminary results of study of merger trees from the Illustris project cosmological simulation. The last item is a part of the wider ongoing effort to study galactic properties related to astrobiology in deeper detail, using both multiple numerical and semi-analytic methods. The exposition in the rest of the paper is as follows. After we enumerate and briefly discuss each of the various primary factors influencing habitability of galaxies in section 2, we consider the secondary/derived/effective properties of galaxies, as well as confusion surrounding them in the recent literature, in section 3. Section 4 presents the method of merger trees as an efficient approach to studying contingent (``historical'') properties of galaxies, while section 5 gives preliminary results of the merger tree analysis on several prototypical subsamples of simulated haloes. Major conclusions are summarized in the concluding section, in which we outline some of the prospects for future research in this novel and exciting area.

\section{2. MAJOR FACTORS INFLUENCING GALACTIC HABITABILITY}

There is a substantial confusion in the literature as to the relative role of various factors obviously or likely influencing galactic habitability. Since it seems clear that there are multiple such factors, the quality of theoretical models is of necessity influenced by the number and manner of accounting for those factors. As we have mentioned above, the original ``rare Earth'' theorists have been mostly motivated by the stage of chemical evolution (i.e., metallicity) in their sweeping claims. Today, we are practically certain that they were wrong in this, although there is still no comprehensive paradigm which could completely substitute for the ``rare Earth'' views. Instead of a comprehensive hierarchy of a priori improbable ``rare'' requirements, we are nowadays facing more complex ``temporal windows'' for particular evolutionary steps to occur or not: ground-breaking study of Chopra and Lineweaver (2016) has demonstrated how contingency creeps in even when all ingredients required by the ``recipe'' for a viable biosphere are present.  This implies that even if all external constraints are satisfied, there is further work on internal, ecological, and macroevolutionary biotic regulation to be done in order to predict a number of biospheres in any sufficiently large cosmological volume.   

In this section we review these external (major) factors which have been either explicated or alluded to in the literature dealing with galactic habitability, while in the next section we shall consider the effective and arguably misleading criteria one sometimes encounters. The factors discussed here are not ordered according to their importance in making the final conclusion about habitability -- which is impossible to do anyway, since we are just starting to weight them in a serious, quantitative manner -- but according to a degree of epistemic certainty which we assign to individual items.

One also needs to keep in mind that even those are quite simplified. The fact that we can establish causal relation between what astrophysicist call ``metallicity'' and our notions of habitability does not make metallicity any less of a simplification for talking about complex network of chemical abundances which do not need to conform to the pattern we see in the standard (Solar) chemical composition or in the Milky Way disk or halo. It has been argued for quite some time that particular quirks of terrestrial biochemistry like the preponderance of phosphorus over sulphur in living beings, or large biochemical role of an extremely rare element like molybdenum over his more abundant (by orders of magnitude!) metallic relatives chromium and nickel (Crick and Orgel 1973), could indicate special properties of the sites of origin of life. To subsume all the richness of biochemistry and its abundance patterns into simple [Fe/H] value is obviously simplistic and serves just as a rule-of-thumb which requires further elaboration on both astrophysical and biochemical sides.

\subsection{2.1 Age}

It is clear that galaxies need to be of certain age to be habitable -- the ``time and chance'' of evolution clearly require that the potential habitats exist for a sufficiently long time before they might be considered habitable. In essence, it is just a restatement of the fact that whatever makes a site habitable is a continuation of the cosmological, astrophysical, chemical, etc. evolutions. And all evolutionary processes without exception require time. In this sense, the point of Gonzalez (2005) that we should talk about ``cosmic habitable age'' in a wider, 4-D spatiotemporal context, is indeed well taken. The build-up of metallicity in the course of chemical evolution requires time, which has first been quantified by Lineweaver (2001). There can simply be no doubt that a certain threshold age of a given galaxy is a necessary prerequisite for any locale, planetary or otherwise, in it to be habitable, and thus that the galaxy as a whole is habitable. A fine representation of this is the large grey area in the upper central part of Figure 3 of Lineweaver, Fenner, and Gibson (2004). 

It is clearly useful at this juncture to recall that the age requirement is a simple consequence of the evolving universe in the standard (``Big Bang'') cosmological model. Irrespectively of its exact form -- we shall return to the current $\lambda$CDM below -- there was an era of the early universe with no galaxies, and consequently no galactic habitability. After the structure formation processes have already been ongoing for some time, first Population III stars appeared and created first metals, setting the background for subsequent dynamical and chemical evolution of galaxies, hence creating necessary conditions for the very notion of galactic habitability to become meaningful. This is confirmed in the numerical simulations of GHZ mentioned above, in which first snapshots of the early disk history show no habitable sites whatsoever. While it is still highly model-dependent, there is a well-defined epoch in our cosmic past when that happens and the very first habitable planets appear.

(This is drastically different in comparison to, say, defunct but historically important steady-state cosmological theory of Bondi, Gold, and Hoyle, which contained galaxies of -- at least in principle -- all possible ages and in which it made no sense to even speak about the average age of galaxies without specifying the volume/sample under consideration. It was exactly the proof that many uniformly spatially galaxies passed through a transient nuclear activity, specifically quasar, phase many billions of years ago without anything similar happening now or in the recent past which contributed to the downfall of the steady-state theory; see Kragh [1996] for more details.)

From the point of view of studying galactic habitability today (i.e., of very recent past, cosmologically speaking), the age requirement seems to be taken for granted. This might, in fact, have some quite practical consequences in light of a couple of recent important developments in the domain of extragalactic SETI. Notably, the G-hat survey for traces and manifestations of higher Kardashev Type civilizations (Wright et al. 2014a,b; Griffith et al. 2015) and the procedure for searching for Type 3 civilizations in external spiral galaxies via Tully-Fisher relationship (Annis 1999; Zackrisson et al. 2015) have probed the universe to large distances. In particular, the most distant objects in the WISE G-hat survey are located at about 100 Mpc, and the most distant spiral galaxy candidate of the Zackrisson et al. (2015) survey is at about 180 Mpc (comoving radial distance).

These extragalactic objects have a non-negligible \textsl{lookback times}, as mandated by relativistic cosmology. Clearly, if the lookback time is sufficiently large, no manifestation of life or intelligence could be detected as a matter of principle. One may instructively think about it putting oneself in a position of an extraterrestrial observer trying to assess habitability of our Earth from a sufficiently large cosmological distance; we can conveniently ignore the extremely high level of sensitivity of the observational equipment and focus just on the spatiotemporal distance. Even in the simplest, Euclidean cosmological model, no observer located at distances larger than about 5,000 ly, or 1,530 pc, could see any indications of human civilization, those at distances larger than about 5 million ly (1.53 Mpc) would not see any hominins at all, those at distances larger than 540 million ly (165 Mpc) would not be able to detect biosignatures of metazoans; those at distances larger than about 4 billion ly (1,230 Mpc) would not see any life at all; finally, observers located at more than 4.5 billion ly (1,380 Mpc) could not have detected the presence of Earth and the Solar System in the first place. Relativistic cosmology, in particular the new post-1998 standard $\lambda$CDM model, changes these figures slightly -- but that is unimportant for the gist of the argument: \textsl{cosmologically recent astrobiological features could be lacking from any extragalactic sample due to large lookback times to the source}. 

This is also an additional motivation to try to elaborate and constrain the role of ages and time intervals in the overall astrobiological landscape. Since the extragalactic sector of the astrobiological and SETI research is of quite recent origin, it is not surprising that these questions have not been posed earlier. Maximal lookback times in our Galaxy are about $10^5$ years, which is relevant to SETI and prevents us from observing young civilizations and our peers, but it does not play a role in habitability as such. After all, Earth was as habitable as it is now $10^5$ years in the past, and with due attention paid to mass extinction episodes in our evolutionary past, anything smaller than about $10^8$ years probably did not involve significant changes in habitability. On the other hand, habitability of galaxies has to take this into account, while adopting the best available timescales from evolutionary biology and palaeontology.   

\subsection{2.2. Chemical Abundances/Metallicity}

Until very recently, considerations of chemical abundances and metallicity have been considered paramount in studying any environment sufficiently different from Earth. As mentioned above, metallicity constrains were the main motivation for introducing GHZ in the first place. Quantification of metallicity as selection effect enabled Lineweaver's (2001) pioneering analysis of the age distribution of terrestrial planets. Concerns about metallicity of exoplanets also led to the erroneous conclusions that early-type galaxies cannot be hospitable to life. 

Another factor which has contributed to relative lack of emphasis on metallicity in recent years is the realization that the neat picture of uniform metallicity gradient in the Milky Way is no longer valid. Instead of a single gradient (e.g., Tadross 2003), we are apparently dealing with a clumpy distribution of chemical enrichment, characterized by discontinuities (Pedicelli et al. 2009; Cheng et al. 2012). In such a situation, it is entirely conceivable that a high-metallicity clump in an otherwise relatively low-metallicity galaxy could harbor life and intelligence. 

Of course, all metallicity concerns apply inside individual galaxies as well. In studying potential habitability in globular clusters, we again see continuity between habitability on intra-galactic vs. habitability on extragalactic spatial scales. The reasons invoked by rare-Earth theorists against early-type galaxies as abodes of life match those traditionally used against globular clusters (e.g., de Juan Ovelar et al. 2012). Conversely, the relaxation of the metallicity requirement in the post-\textsl{Kepler} period led to increased optimism regarding astrobiology of globular clusters (Di Stefano and Ray 2016). This is based upon conjectural ``sweet spots'' in which adverse causative factors, such as low metallicity, are outbalanced by those advantageous for the emergence of life, such as longevity or low irradiation rate. The same lesson has now been, as sketched in Section 1, applied to all external galaxies.

\subsection{2.3. Stellar Explosion Rates}

Intermittent sources of ionizing photons, as well as still controversial jets of cosmic ray particles, such as supernovae, GRBs, or magnetar explosions can adversely influence habitability through a host of in general poorly understood ecological processes. Those usually invoked include atmospheric chemistry changes, accompanied by ozone layer destruction, drastic albedo changes, acidification of surface waters, etc., as well as the direct damage upon unshielded biological polymers such as DNA or protein chains. All of these are to some degree necessarily geocentric (for instance, it is reasonable to assume -- although no specific studies have been performed so far -- that hypothetical subglacial ecosystems on Europa would be significantly more resistant to this kind of perturbations due to the thick ice shielding). Lacking firm evidence about radically different biospheres, however, it is reasonable to use what has originally been aimed at estimating the risk of stellar explosions to Earth and its biosphere (e.g.,  Crutzen and Br{\" u}hl 1996). In this limit, it is indubitable that such outbursts of both high-energy photons and cosmic ray particles are capable of either destroying or severely damaging biospheres on planets within a ``lethal zone'' of the event. (Here we consider only stellar explosions, while leaving the outbursts of galactic nuclei for the next subsection.)

How big is that ``lethal zone'' for various kinds of intermittent sources is the central issue here, which transforms into the size of the irradiated volume via statistics of stellar populations. For classic supernovae of Type Ia or Type II it has been estimated as being on the order of 10 pc, most frequently between 10 and 30 pc (e.g., Ellis and Schramm 1995; Gehrels et al. 2003; Melott and Thomas 2011), with newer studies favouring lower values. For GRBs or its progenitor ``hypernovae'' similar to $\eta$ Carinae (nominal distance 2.3 kpc from the Solar System), Scalo and Wheeler (2002) give large radii of lethal gamma-ray fluxes as 1.4 kpc for prokaryotes and 14 kpc for eukaryotes. The latter would span most or all of GHZ, depending on the degree of collimation. In contrast, Thomas et al. (2008) conclude that the lethal radius of effect of the $\eta$ Carinae type event is fairly small: only about 0.3 pc for the extinction-level ozone layer disruption. (Possibly an order of magnitude more for other forms of radiative disruption.)

The key shortcut here is that the rate of stellar explosions is proportional to the star formation rate (e.g., Hopkins and Beacom 2006; Li and Zhang 2015). \, This is certainly justified for Type II supernovae and hypernovae/longer-GRB-progenitors. Type Ia supernovae, and shorter GRBs originating in neutron star mergers can, however, occur in any type of galaxy, including those with little or no star formation at $z = 0$ (although it does not mean that their rates depend on stellar mass of the galaxy only; cf. Sullivan et al. 2006). This likely applies to magnetar bursts, whose general rate per unit mass of galactic volume (and especially how the rate changes with cosmological time) is unknown. Careful correction for all these effects in the context of galactic habitability has not been done so far. 

While electromagnetic radiation coming from stellar explosions is known fairly well, another potentially disruptive conundrum is the fluence of cosmic ray particles accelerated in supernovae and GRBs. The whole spectrum of views on this issue is present in the literature (for more than half a century already, cf. Laster, Tucker, and Terry 1968), from entirely neglecting the cosmic ray production in these events, to the view that stellar explosions -- GRBs in particular -- are the main source of high energy cosmic rays both within host galaxies and in general, and hence are astrobiologically relevant (Dar, Laor, and Shaviv 1998; Dermer and Holmes 2005; Ferrari and Szuszkiewicz 2009; Erlykin and Wolfendale 2010; Dartnell 2011 and references therein; Marinho, Paulucci, and Galante 2014). Even a staunchly optimistic study such as the one of Thomas et al. (2008) which explicitly does not cover cosmic-ray production in its assessment of the threat of $\eta$ Carinae admits that there is substantial uncertainty on this issue. All this is somewhat ironic in light of the fact that the role of cosmic rays in mutagenesis was historically one of the very first things ever established about genetic information and mutations (e.g., Babcock and Collins 1929); this has been confirmed by numerous studies in modern molecular biology and genetics, some performed within an explicitly astrobiological context (Moeller et al. 2010). Even conservative evolutionists such as George G. Simpson admitted that cosmic rays likely play an important role in macroevolution as well (Simpson 1968). Since the relevance of cosmic rays \textsl{sui generis} for terrestrial biology was never seriously in doubt, it is somewhat surprising to find relative paucity of research and specific results on their wider astrobiological impact.

Nothing illustrates better the dramatic confusion reigning in this domain than the already noted fact that conclusions of the several existing studies of the habitability of galaxies dramatically differ on this point. While Dayal et al. (2015) regard stellar explosions as the paramount determining factor, Gobat and Hong (2016) find them negligible, except in the limit of the largest total mass. Stanway et al. (2018) consider irradiations by stellar explosions as the dominant factor, and the metallicity threshold as only subsidiary. They find no significant dependence of the irradiated mass fraction on the stellar mass of the system, in contrast to Gobat and Hong. Clearly, much further work on this sub-topic is required in order to clarify this confusion.

\subsection{2.4. The Nuclear Engine}

One way in which our understanding of galactic habitability, when based \textsl{entirely} on our Milky Way experience, might be seriously and systematically biased is the neglect of the effects of central nuclear engine. Nuclear engines of galaxies usually imply central supermassive black hole which swallows matter in shorter or longer accretion episodes, emitting energy in practically all parts of the electromagnetic spectrum and being seen as active galactic nuclei. The intermittent episodes are separated by long intervals of quiescence corresponding to ``normal'' galactic cores. We can thus regard the nuclear outbursts in the active phase as a special case of stellar explosions: the one occurring at the same spatial location, last longer ($\sim 10^7$ years), and is of much larger total output. Local ecological effects adverse to habitability are expected to be the same as for planets in the ``lethal zone'' of a supernova or a GRB. 

Clarke (1981) was the first to consider the effects of galactic nuclear activity on our search for life and intelligence elsewhere; as it happens, science fiction was there first: \textsl{The Inferno}, novel written by Sir Fred Hoyle and his son Geoffrey, depicts possible consequences of such an outburst in the Milky Way nucleus for Earth (Hoyle and Hoyle 1973).\footnote{For a more up-to-date literary description of the ecological effects of both GRBs and nuclear outbursts, see Egan (1997).} In a brief and qualitative paper, Clarke has suggested the large-scale galactic environment plays decisive role in emergence or not of extraterrestrial intelligence, hence in a galaxy being viable SETI target or else. He based his view on the idea, somewhat fashionable in the late 1970s, that the nucleus of the Milky Way can undergo Seyfert-like recurrent bursts of activity (e.g., Sanders and Prendergast 1974; van Bueren 1978; Clube 1978). Although the physics employed (or implied) by Clarke is outdated, the central point is still quite interesting from the astrobiological point of view: since nuclear outbursts were provably more frequent in early epochs of galactic history, we may not be so surprised that the universe becomes more hospitable for life with passing of cosmic time.

The study of Balbi and Tombesi (2017) investigates atmospheric loss and biological damage on terrestrial planets in the inner parts of the Milky Way (and, by analogy, any large galaxy). The nuclear source of our Galaxy, Sgr A*, is currently in a quiescent phase, but during the epochs of peak activity it could cause total atmospheric loss within a kiloparsec and arguably suppress habitability in the inner several kiloparsecs. The above-discussed study of Gobat and Hong (2016) uses the following analytical ansatz for active nuclei:
\begin{equation}
\log{R_\mathrm{AGN}}=-6.1+0.7\log{M_*},
\end{equation}
where the lethal radius $R_\mathrm{AGN}$ is in kpc and the stellar mass of the galaxy $M_*$ is given in Solar masses. While this gives plausible order-of-magnitude estimates, there are many relevant effects which have not been captured by it, such as the extinction and shielding in the disk, and especially the sustained nature of the AGN emission when compared with the stellar explosions. Again, just as in the case of stellar explosions, the issue of production of biologically relevant cosmic-ray particles in AGN outbursts is largely open. 

Leaving aside the question of particularities (or the lack) of the Milky Way nuclear engine, it is clear that the importance of this mechanism for irradiation will vary much from galaxy to galaxy. Giant galaxies like the Milky Way or M31 have more massive central black holes and are expected to have stronger outbursts of nuclear activity -- in the same time, their GHZs are much wider and, crucially, they possess much more interstellar matter, creating larger optical depths for ionizing radiation and (possibly) cosmic ray particles. Dusty giant molecular clouds would thus present strong shielding against nuclear outbursts, at least for the outer parts of GHZ and for planetary systems located close to the plane of the disk. Also, the satellites of such active galaxies might find themselves right within the AGN jet angle (for analysis of a few other examples of galactic interaction see also Section 3), with adverse consequences for habitability. 

On the other hand, dwarf galaxies have more modest nuclear engines and there are indications from numerical simulations that accretion onto central supermassive black hole is both qualitatively and quantitatively different in dwarf subhaloes (Micic, Martinovi{\' c}, and Sinha 2016). Prolonged, low-level accretion might be the preferred mode of nuclear activity in most dwarfs, creating persistently higher ionizing background all over the place; astrobiological consequences of such an environment have not been studied yet. 

\subsection{2.5. Dynamical Stability}

Dynamical stability of planetary systems at long timescales characteristic for biological evolution has been seen in Gonzalez et al. (2001) as one of the reasons for excising the innermost part of the Galaxy from GHZ. The reasoning is fairly simple: since continuous habitability requires time, by definition, catastrophic changes of planetary orbits brought about by very close passes of nearby stars limit habitability in the given environment. Since the rate of such stellar encounters is given as $<\!\!nv\sigma\!\!>$, where $n$ is the stellar density, $v$ relative stellar velocity and $\sigma$ cross section for the close pass having adverse astrobiological consequences, we are really considering only regions with very high stellar density. Laughlin and Adams (2000) estimate that truly disruptive close passes occur with $\left\langle \sigma \right\rangle \sim 100$ AU$^2$, although it might be too conservative estimate, since even larger values are likely to lead to built-up eccentricity of the planets in CHZ and cause strong perturbations of planetary climate. Obviously, whatever specific threshold we use for defining what is adverse for habitability, a degree of subjectivity in this parameter will remain. On the other hand, whatever the most appropriate value for this cross-section might be, once established, it stays constant or nearly constant everywhere and at all times. As far as relative velocities of stars go, the values for $v$ do not vary for more than a factor of a few either within the Milky Way (roughly $30-120$ km s$^{-1}$) or in external galaxies. Therefore, the controlling parameter is the average stellar density.  

Note that this factor does not only play role in the innermost parts of \textsl{all} galaxies (since all kinds of galaxies studied so far possess the central maximum of stellar density), but also in other structures which present local enhancements of stellar density. In particular, this applies to the globular cluster environments, whose habitability has recently been a subject of some controversy, as mentioned above. This also strengthens the case for strong continuity between dwarf early-type galaxies and globular cluster environments. Another interesting component of external galaxies which have not been studied in this context so far are polar rings of galaxies such as NGC 2685 (the Helix Galaxy), NGC 660, or NGC 5128 (Centaurus A).

\subsection{2.6. Wild Cards?}

One should be very careful at this point to avoid claiming that the items enumerated above are indeed all and everything influencing (or even determining) galactic habitability. It is reasonable to assume that there are ``wild card'' factors in play as well, we are currently not aware of. There is an important and instructive analogy in our understanding of habitability of planets -- and indeed the habitability of our Earth in its early epochs. The already mentioned study of Chopra and Lineweaver (2016) emphasizes that \textsl{continuous} habitability depends on a number of complex, nonlinear feedback loops. In such loops, a small forcing at a particular early epoch of planetary/biospheric evolution could cause the runaway response, unbalance the system, and ultimately make the planet uninhabitable. Since those early forcings are small, it is easy to miss them, especially since there is no ``unified theory'' of the terrestrial planets' evolution. Therefore, prebiotic or early biospheric environments could be much more fragile than we naively think -- and therefore we need to look out for possible ``wild card'' influences.

An example of such a ``wild card'' influence which has been intuited, but not been studied in detail so far, is the effect of a planetary system with at least one habitable planet encountering an interstellar cloud of gas and dust. Such occurrences have to be frequent during the lifetime of habitable planets, but the effects are necessarily quite varied since both properties of clouds such as density, gas-to-dust ratio, opacity, etc. and parameters such as relative velocity or geometry of the encounter may vary wildly. That encounters with dense clouds could lead to strong perturbations of Earth's climate has been proposed by Begelman and Rees (1976). The idea has been elaborated by Pavlov et al. (2005) for Galactic giant molecular clouds with densities of $10^3$ cm$^{-3}$ or greater. Their results suggest that extreme glaciations, known as the ``Snowball Earth'' episodes (Hoffman et al. 1998) could be triggered by the accumulation of stratospheric dust and the consequent ice-albedo feedback. At least two ``Snowball Earths'' confirmed during the Precambrian were extreme global cataclysms, which might have led to the extinction of life or at least drastic reduction in its complexity. Also, the relationship between giant molecular clouds and close core-collapse supernovae has been recently studied (Kokaia and Davies 2019). For possible link to controversial periodicities in terrestrial mass extinction episodes, see Filipovic et al. (2013). Much further work will be necessary to assess the extent such occurrences may be general and whether the encounters influence habitability at large scales. One interesting aspect of this possible mechanism is that it is clearly selective on the level of galaxies: only late-type (and possibly some irregular) galaxies are sufficiently gas- and dust-rich for this mechanism to operate. Conversely, relative habitability of early-type galaxies could be enhanced in this manner. 

\section{3. MORPHOLOGICAL TYPE AND OTHER DERIVED CRITERIA}

Properties such as metallicity or star-formation rate have been, until recent advent of large catalogues like SDSS, quite difficult and expensive to establish for any large sample of extragalactic sources. Hence the attempts to take shortcuts and reach quick and rough conclusions about galactic habitability. It is certainly necessary to discuss some of other criteria which have been thrown around in debates about galactic habitability, for dispelling some of the prevailing confusions if nothing else. 

In a sense, studying habitability of galaxies faces problems similar to the age-old puzzles surrounding biogeography of continents on Earth: each of them contains too complex and intricate web of habitats and ecosystems, they have rough similarities and dissimilarities, and yet they are too few to allow for meaningful statistical treatment.\footnote{In a somewhat eerie coincidence, biogeography on the scale of continents has been also pioneered by Alfred R. Wallace in his great monograph \textsl{The Geographical Distribution of Animals} (Wallace 1876; for a modern overview, see e.g., Humphries and Parenti 1999).} There are many galaxies, but only a small number could be observed and modelled in detail sufficient for astrobiological purposes. For most purposes, a very coarse statistical treatment of their properties is possible -- the examples of this coarse-graining are gross simplifications inherent in using terms gas-phase and stellar metallicity, as mentioned above. The coarse-graining here is not dissimilar to the usage of this expression in statistical mechanics, since it allows us to use the theoretical framework to separate ``really important'' macroscopic changes in the system from purely random, chaotic microscopic changes which leave the macroscopic state unchanged. By analogy, we wish to study those evolutionary processes which significantly impact galactic habitability in contrast to those myriads of events which are ``just history''. 

As stated in the introduction, the position of the ``rare Earth'' theorists was that the early-type galaxies, as well as most dwarfs, are not habitable due to low metallicity and consequent suppression of the terrestrial planet formation. Apart from gradual understanding that the early correlation of exoplanets with metallicity was mainly due to observation selection, we have also come to understand that metallicity in dwarfs -- those which are observable in more detail, mainly in the Local Group (McConnachie 2012) -- need not necessarily be as low as it was previously assumed (Peeples, Pogge, and Stanek 2008). This conclusion has some independent theoretical support as well (e.g., Kirby et al. 2013). We have seen that the new wave of thinking, inaugurated by the work of Dayal et al. (2015), implies that the early-type galaxies are indeed the most habitable kind of galaxies. There are tantalizing indications that this applies to some dwarfs as well (Stojkovi{\' c} et al. 2019). All in all, it does seem that the morphological type \textsl{per se} is a poor indicator of the habitability of any particular galaxy.

This has another, more fundamental consequence. For all the reasons outlined above, it is very difficult to be certain whether some aspects of the evolution of galaxies are systematic, lawful occurrences dictated by general dynamics at the galactic spatiotemporal scales, or historical contingencies based on particular details of each individual case.\footnote{This is perfectly analogous to the convergence vs. contingency debate in evolutionary biology, where the matter of contention and often fierce debate is whether particular trait of a species or a lineage is either (i) a consequence of the general evolutionary mechanisms like natural selection, which will tend to give regular, lawful outcomes; or (ii) a consequence of essentially unrepeatable and unpredictable historical contingency. For classical popular expositions of the extremes of these two views see Gould (1989) and Conway Morris (2003); for a somewhat refined view with important consequences for astrobiology, see Vermeij (2006).} Individual histories of galaxies could be sufficiently peculiar due to historical contingency to invalidate any sweeping conclusion based on simplified criteria such as morphological type. The case of M32, the bright dwarf satellite of M31 (the Andromeda Galaxy) is very instructive in this respect. M32 belongs to the compact dwarf ellipticals, being officially classified as cE2 in NED, characterized by high surface brightness and very small effective radius ($R_\mathrm{e}=100$ pc; Mateo 1998). There have been many hints that M32 is peculiar due to its particular environment; of course, the fact that this environment consists of M31 and the whole of the Local Group makes it special \textsl{only for us}, not special in general. We should resist anthropocentrism -- or the Milky-Way-centrism at it happens here -- and all the pitfalls in generalizing from our very limited experience and our own viewpoint in this regard. M32 is simply the easiest cE galaxy to observe and make any reasonably detailed models. Bekki et al. (2001) argue that gravitational tidal forces exerted by M31 have driven the interstellar gas in M32 inward and triggered a burst of star formation, resulting in both high central stellar density and relatively high metallicity observed today. As an alternative, M32 and other compact ellipticals might emerge as a consequence of tidal stripping of giant spirals (e.g., Martinovi{\' c} and Micic 2017) and this in particular might apply to M32 as has been recently suggested by D'Souza and Bell (2018). According to this idea, today's M32 is essentially the bulge of a previously existing progenitor object M32p, whose bigger part was stripped by and absorbed by M31 as late as 2 Gyr ago.

It is extremely instructive to consider consequences of such an evolutionary trajectory of habitability. If D'Souza and Bell are correct, the transformation of M32p into today's M32 occurred late enough so that there was essentially the same temporal window of high metallicity among M32p's stars as is the case for the Milky Way. The disruption provided M31 with an influx of habitable planets, while the degree of habitability present in the core regions of M32p should stay essentially the same. The original GHZ in M32p was disrupted, but it is not obvious that the cumulative contribution of M32p to the total habitability of the Local Group was diminished. In contrast, in the ``threshing'' scenario of Bekki et al. (2001), the starburst triggered by tidal compression was likely an astrobiological ``reset'' (cf. Vukoti{\' c} and {\' C}irkovi{\' c} 2008), interrupting continuous habitability. Of course, in that scenario we should still be interested in the present-day habitability of M32 is the starburst occurred sufficiently long ago and thus there was time for terrestrial planets to recuperate.

The main lesson is thus rather clear: different evolutionary histories of the very same object could imply quite different habitabilities. We need specific methodology -- to be elaborated in the next section -- which is focused on evolutionary histories of galaxies and the environmental impact on those histories, in order to discriminate between ``regular'' histories of galaxies and ``interlopers'' and ``freaks'' such as M32. We need the way to assign statistical weights to different cases in order to see how important they are for the overall average habitability calculation. Note that this would be true even if we abstract away the assumption that the origination of life and its spreading are highly nonlinear processes where even a small increase in habitability might result in large increase in \textsl{detectability} in terms of both biosignatures and technosignatures (Linguam and Loeb 2018). 

Of course, there is a serious problem of observation-selection effects. Per analogy with the reasoning of Haqq-Misra, Kopparapu, and Wolf (2018) in connection to stars having habitable planets, we might wish to ask why we find ourselves in a giant spiral galaxy instead of an early-type galaxy and/or a dwarf. If knew nothing whatsoever about habitability of galaxies, finding ourselves in a large spiral such as the Milky Way would offer a probabilistic support to the original rare-Earth idea that large spirals are only galaxies with nonzero habitability. If the distribution of habitability over morphological types is more complex, however, the degree of support is much more contentious and it requires a careful Bayesian analysis. In contrast to the stellar case, we do not know the final states of galaxy evolution, so it is not appropriate to weight likelihood by temporal duration ({\' C}irkovi{\' c} and Balbi 2019, submitted). One strategy is, obviously, to reject the anthropic reasoning and accept that the Milky Way is an exceptional fluke, in particular by being too quiet (Hammer et al. 2007) in terms of merger- and star-formation histories (cf.\/ Yin et al. 2009). The discovery of so-called Fermi bubbles (Su, Slatyer, and Finkbeiner 2010) casts doubt on this conclusion, however. Even if the conclusions of Hammer et al. (2007) are correct, the subset of $(7 \pm 1)\%$ spiral galaxies found to be similar to the Galaxy in a sample of well-studied local systems is not really that minuscule a fraction that anthropic reasoning should trouble us much; it is incomparable with other known fine-tunings where the observed values fall within a fraction of the parameter space equal to $10^{-5}$ or less (Hogan 2000). 

Other approaches are possible, however; it is possible that further research will indicate that there is a temporal cut-off to the habitability of galaxies, perhaps through insofar neglected dynamical aspects of stellar migrations within a galaxy of particular type (see the discussion in Vukoti{\' c} et al. 2016). Alternatively, it is possible that our astrobiological view of the local universe is skewed too much by particular properties of the Local Group, including both the Milky Way and the specific dwarfs mentioned above. So far, there are no observational indications to the extent that the Local Group is indeed typical in a larger set of similar small galaxy groups at $z = 0$. Finally, we may entertain the idea that it is not habitability in general, but only the reference class of observers like us, which has an ending in a finite future epoch and therefore has a finite and relatively small measure.

\section{4. MERGER TREE EXPLORATION}

The new standard $\lambda$CDM cosmological model has proved to be the most successful cosmological paradigm so far (e.g., Turner 2018). Large-scale numerical simulations performed within the $\lambda$CDM framework have become standard and very flexible tools for investigating structure formation, showing high degree of congruence with observations. By modelling the discretized matter density field (both CDM particles and baryons), these N-body simulations solve for the dynamical evolution of particles under the influence of their mutual gravity and track the formation and evolution of gravitationally bound condensations of dark matter known as halos. The prevailing scenario of galaxy formation predicts that luminous galaxies form within these dark matter halos and that their subsequent evolution is to a large degree shaped by the growth of their host halos. One of the key processes by which this growth occur is merging of haloes and, consequently, the merging of luminous galaxies. The timing, mass ratios and other parameters of these mergers are key for understanding of observable properties of galaxies, such as morphology, stellar mass, star formation rate, colour indices, (in)activity of the nuclear engine, etc. (Benson 2010). Therefore, it is of great interest for investigating the evolution of galaxies to be able to ``distil'' the information about these mergers from the treasure troves of information contained in the large-scale cosmological simulations of structure formation.  

Since $\lambda$CDM establishes strong intrinsic link between visible baryonic matter and dark haloes of individual galaxies, in order to study mergers we need procedures which both (i) identify haloes at any given simulation epoch (``snapshot''), and (ii) track them between different simulation epochs. Historically, the task of identifying haloes has followed from the early Press-Schechter work on evolution of spherical overdensities in the expanding universe (Press and Schechter 1974), although other procedures (``halo finders'') have been subsequently devised. It is the second part of the problem -- connecting different snapshots -- which has proved a tougher nut to crack until mid-1990s and which gave the name ``merger tree'' or ``merger tree building'' to the whole enterprise. The method originates with the seminal paper of Lacey and Cole (1993), who constructed the first algorithm (``tree builder'') based on a semi-analytic model that tracks genealogy of haloes from progenitors to descendants over snapshot outputs of the simulation. Modern reviews of the approach can be found in Zhang, Fakhouri, and Ma (2008); Fakhouri and Ma (2008); Jiang and van den Bosch (2014) and Poulton et al. (2018).

From the astrobiological point of view, the merger tree approach, pioneered by Stanway et al. (2018), has the following major merits in dealing with galactic habitability:
\vspace*{0.5cm}
\item{1.} The approach takes into account complex interactions of galaxies with their heterogeneous environment, something which \textsl{ab initio} fully numerical cosmological simulations cannot do.
\item{2.} Very different kinds of descendant galaxies could be studied by harvesting data from the same simulation outputs.
\item{3.} It offers insight into the vexing problem of distinguishing which properties of contemporary or ``late epoch'' or ``local universe'' galaxies are intrinsic product of their evolution in contrast to those properties which are contingent upon peculiarities of the environment. 
\vspace*{0.5cm}

Of course, a single merger tree could hardly perform the required tasks, so it is often necessary to study a larger ensemble. However, taking into account vastly increased amount of data available from several large, public-domain numerical simulation databases (Millennium Run, Illustris Project, Bolshoi-Planck, EAGLE, etc.), this is not a very daunting task. Also, since harvesting the necessary data for building merger trees is done by semi-analytic procedures, the sheer volume of computing is relatively small and the method does not require dedicated supercomputing capabilities. 

Point \#3 is especially important from the point of view of correction for ubiquitous observation selection effects. In dealing with galactic habitability, we always face the possibility of an anthropic bias related to our being born and living in the particular galaxy, the Milky Way, and its unique surroundings, which might be atypical in several important respects \textsl{without that atypicality being obvious} or, indeed, readily observable at present. For instance, we cannot at present be certain to what extent is the Local Group a typical small group of galaxies, in particular as related to its dwarf-galaxy contents. We have no particular reason to believe it is an outlier, so we are justified in cautiously applying the Copernican principle that it is typical within the relevant reference class (similar small galaxy groups), but only further observational data will offer definite judgement on that. Until that moment, we should be on guard against generalizing particular properties of the Local Group galaxies which are clearly important for habitability -- such as relatively high metallicity of dwarfs such as M32 or the Fornax dwarf -- to similar galaxies elsewhere. This is also related to the concept of \textsl{continuous habitability} as discussed above: two galaxies might look superficially similar today, but have sufficiently different evolutionary histories so that one has been continuously habitable in, say, $25\%$ of its stellar mass for the last $5$ Gyr, while in the other the corresponding fraction is $1\%$ or less. It is exactly for this purpose that the merger tree approach offers the best solution -- it may point out what is truly universal in galactic habitability and separate it from the quirks and contingencies of history. 

\section{5. RESULTS}

To illustrate the plausibility and functional usefulness of the method of merger trees, we present, following Stanway et al. (2018), preliminary results from the analysis of a sample of subhaloes from the Illustris Project. The Illustris Project (Vogelsberger et al. 2014; Nelson et al. 2015) is a cosmological hydrodynamical simulation using the moving-mesh code Arepo (Vogelsberger et al. 2012), which includes a comprehensive set of physical models needed for closely following formation and evolution of galaxies. Spatial volume included in the simulation is $106.5^3$ Mpc$^3$. There are six runs of Illustris simulation: three with full baryonic physics models and three with CDM only. While we have used the baryonic physics models, it is worth mentioning that the physics used in the simulation is strictly consistent only for subhaloes with the total mass larger than about $10^9$ Solar masses. 

Nevertheless, the Illustris Project has several advantages over the Millennium Simulation used by Stanway et al., notably higher mass resolution, improved SFR feedback and other aspects of baryonic physics, and a larger number of snapshots available. The galaxy mass-metallicity relationship seen in the local universe (the ``fundamental plane'' discussed in Dayal et al. 2015) is reproduced naturally from cosmological evolution. For comparative analysis of merger trees extracted from Illustris see Rodriguez-Gomez et al. (2015). More detailed analysis of the Illustris merger trees will be given in the course of the future work; for the moment, we present preliminary findings on the selected parameters of a set of subhaloes (haloes which are not at the top of the hierarchy, i.e., lying within one or more other haloes) which correspond to the parameter identified by Dayal et al. (2015) as key to the galactic habitability: stellar mass ($M_*$), stellar and gaseous metallicity ($Z_*$ and $Z\mathrm{g}$), and star-forming rate (SFR). {\bf The Illustris simulation outputs the metallicity as the ratio of the total gas(stellar) mass of all simulated elements heavier than He to the total gas(stellar) mass (figures 2, 4 and 5) .}

We have chosen a sample of {\bf 908} descendant galaxies (subhaloes) for the purpose of this precursor study. The important signposts of their evolution are shown in the figures. In particular, the total (CDM + baryonic) mass of a typical massive descendant subhalo is shown in Figure 1, showing changes which are typical for objects in large-scale N-body simulations. This would be an analogue to a large spiral galaxy, similar to the Milky Way or M31 at present epoch.  

}
%\end{multicols}

\includegraphics[width=\textwidth, keepaspectratio]{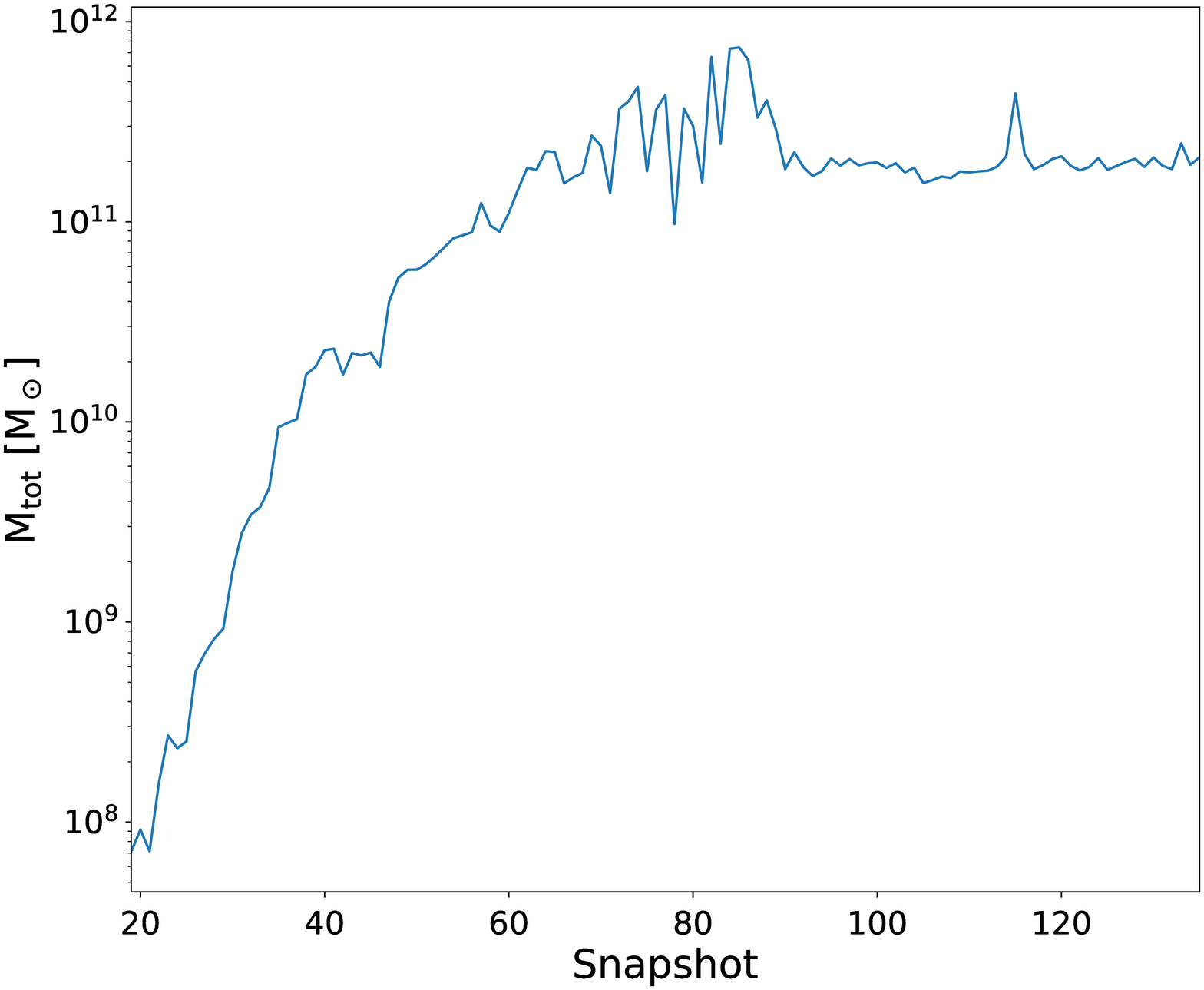}

\figurecaption{1.}{The change and growth of the total mass of a typical selected simulated galaxy from Illustris during its history. The change in the total mass from the first progenitor of the given galaxy ({\bf 19}th snapshot; $\mathbf{z = 18.79}$) to the current epoch (135th snapshot; $z = 0$) is due to the numerous mergers that the galaxy goes through during its life, where the downwards trend followed by a sudden increase in total mass is the characteristic of the merger event (Rodriguez-Gomez et al. 2015).}

%\begin{multicols}{2}
{

The average stellar metallicity and the normalized SFR for the same prototypical subhalo are shown in Figure 2 for the same range of snapshots ($27-135$). It is evident that in its early history, the star formation rate was rapidly increasing, and parallel with the growth of the SFR, the growth of stellar metallicity is evident. A very steep increase in early metallicity is due to the assumed physics of the hypothetical Population III stars, which is built in the simulation and should be taken with reserve. Subsequent minor declines in metallicity visible in the diagram are probably due to the inflow of low-metallicity gas during minor mergers and other processes (e.g., galactic wind ``blowout'' could result in lowering mean metallicity of the subsequent generation of stars).
}
%\end{multicols}

\centerline{\includegraphics[width=\textwidth, keepaspectratio]{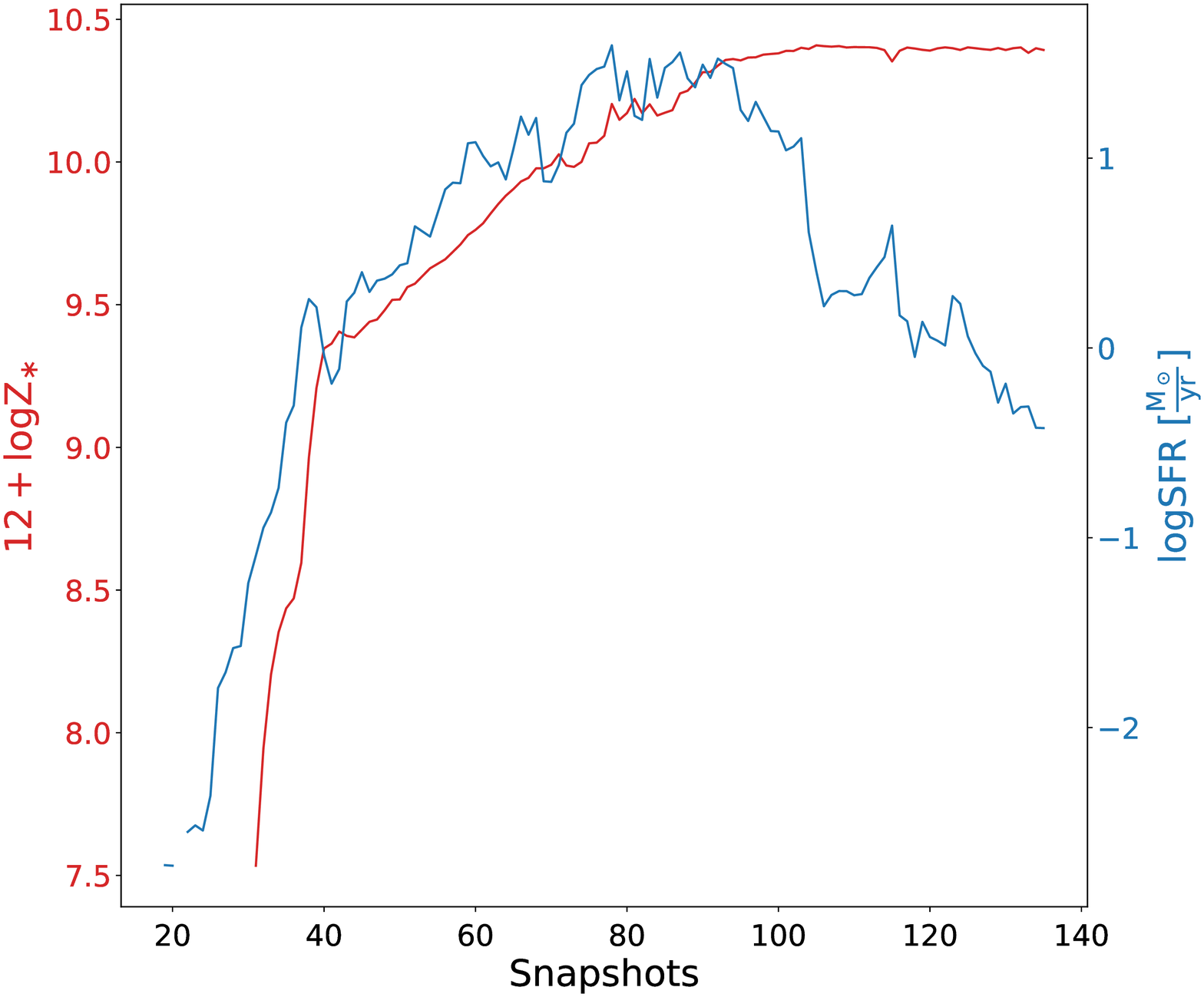}}

\figurecaption{2.}{The star formation rate (blue line, blue ticks on the right-hand side of the diagram) and stellar metallicity  (red line, red ticks on the left-hand side of the diagram) of the selected galaxy during its history.}

\centerline{\includegraphics[width=\textwidth, keepaspectratio]{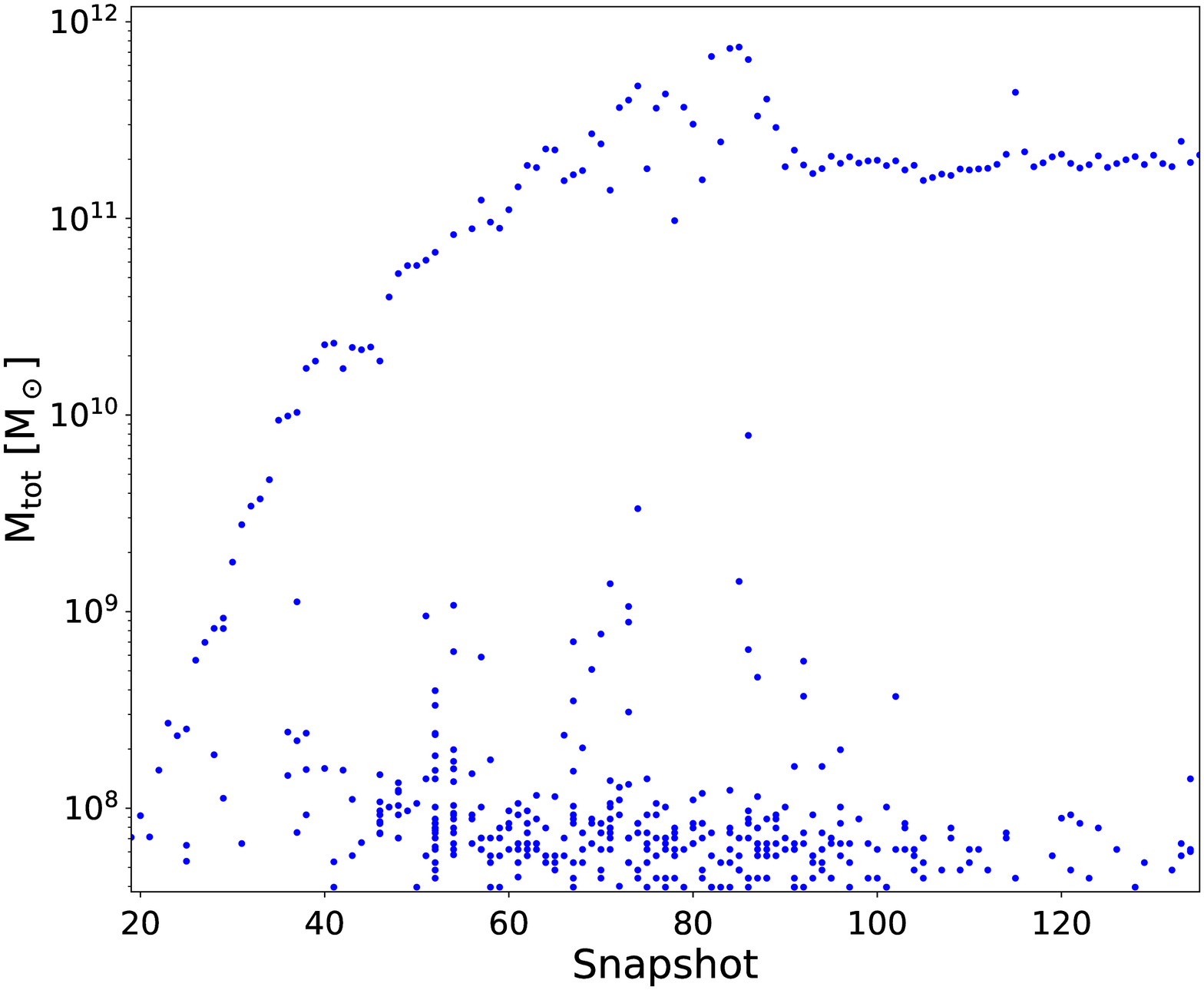}}

\figurecaption{3.}{The image shows all of the galaxies which the selected galaxy collided with during its history. The galaxy with the highest mass history in every snapshot is the one whose history is followed, and in every snapshot that is usually galaxy with the highest total mass (the ``central''). Other galaxies within the same snapshot are secondary participants in the mergers and usually have much lower total mass (``satellites'').}

%\begin{multicols}{2}
{

The complex nature of the merger network is demonstrated in Figure 3, showing mass history of chosen galaxies. By definition, a merger takes place when a galaxy has more than one progenitor. Direct progenitors at snapshot n are usually identified in the previous snapshot $n-1$, although there might be exceptions in which they are found in the $n-2$ snapshot (see Rodriguez-Gomez et al. 2015 for more details). In the ideal case, this implies that if a chosen galaxy has Np progenitors, there have been $N_\mathrm{p}-1$ mergers in its history; in practice, this is dependent on the temporal/redshift resolution of the simulation (Fakhouri and Ma 2008). In this respect, the Illustris Project is superior in comparison to the previous simulations, having the effective redshift resolution $\Delta z \approx 0.01$ at low redshifts, which is fine enough for vast majority of mergers to be captured and counted. 
}
%\end{multicols}

\centerline{\includegraphics[width=\textwidth, keepaspectratio]{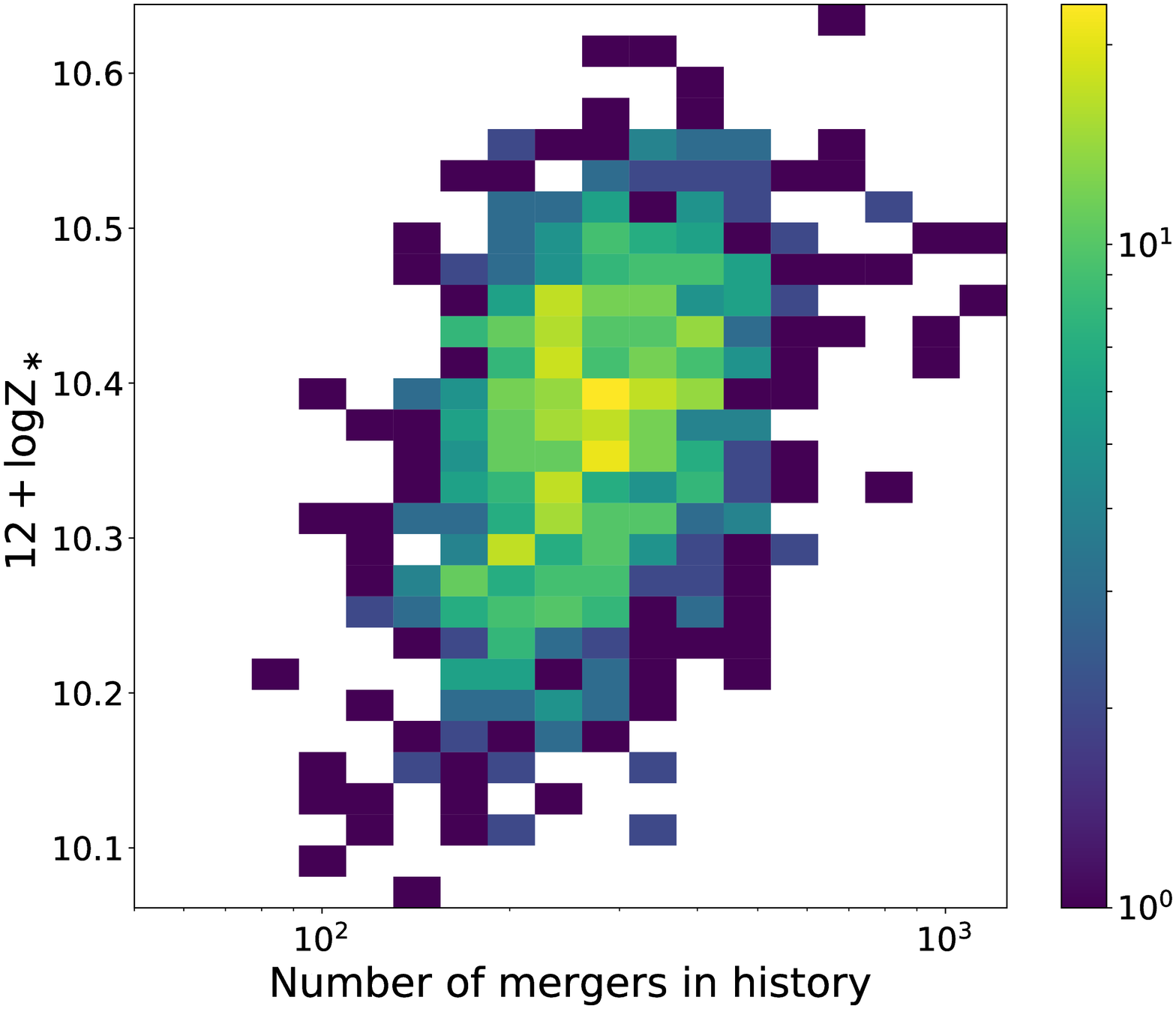}}

\figurecaption{4.}{The sample shown in this image is made up of {\bf 908} galaxies of total mass in range of $\mathbf{[8 \times 10^{11} - 2 \times 10^{12}}]$ M$_\odot$. The color bar shows the number of galaxies in certain ranges of star metallicity in the logarithmic, non-zero form (y-axis) and the total number of mergers in their history (x-axis). The histogram shows a clear trend of increasing stellar metallicity with increasing the number of mergers in the galactic history.}

\centerline{\includegraphics[width=\textwidth, keepaspectratio]{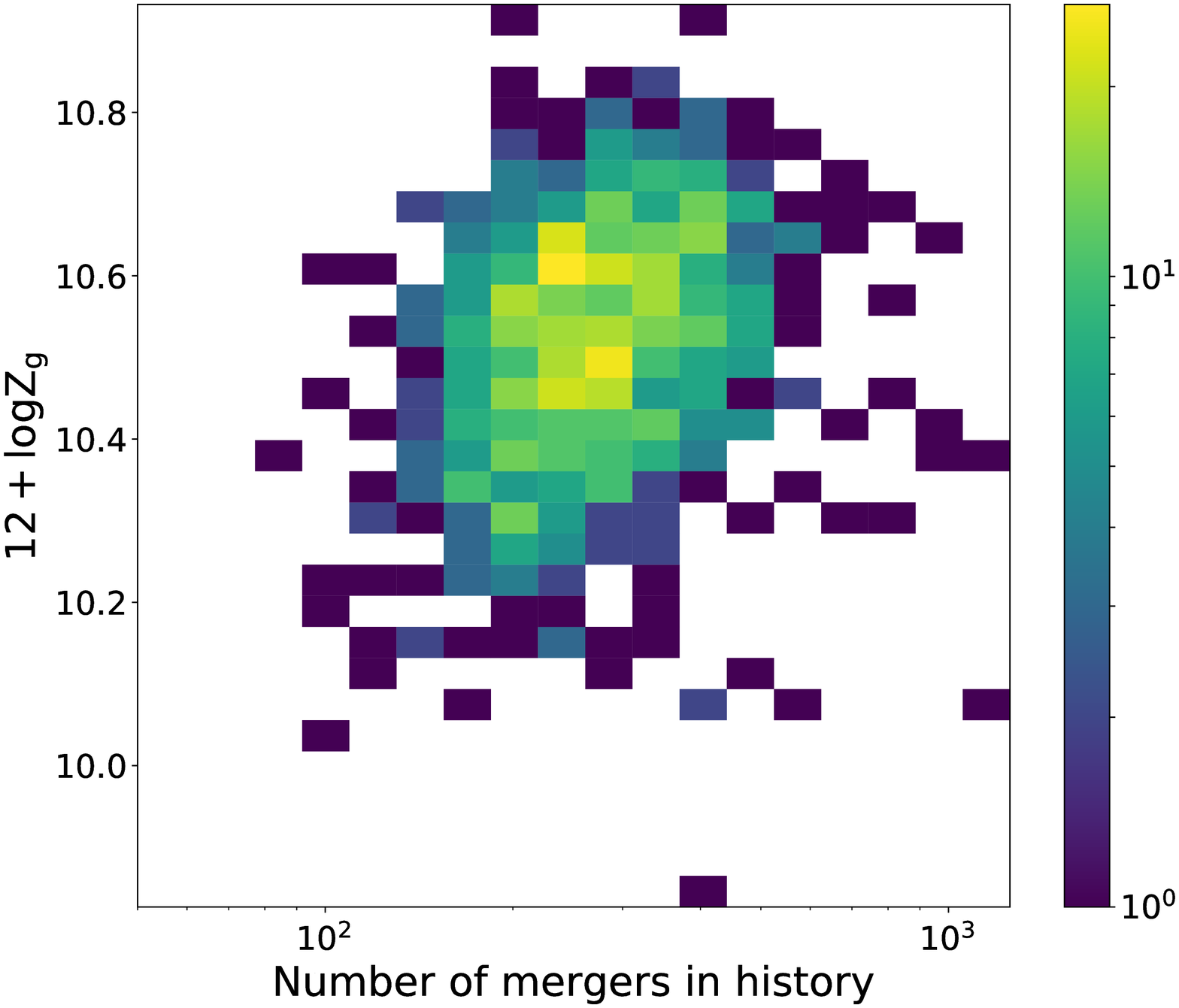}}

\figurecaption{5.}{The same sample as in Figure 4, now stratified by the gas-phase metallicity (y-axis) and the total number of mergers in their history (x-axis). The trend of the increased metallicity with a higher number of mergers {\bf is present similar as in Fig. 4.}}

%\begin{multicols}{2}
{

Comparing results in Figure 4 and Figure 5 it is clear that the increase of the mean metallicity of the stellar population {\bf appears to have a slightly better correlation} with the number of mergers occurring in a galaxy's history {\bf when compared to} the analogue increase for the interstellar gas. This is intuitively acceptable, since most of the mergers have occurred with smaller galaxies and at high redshifts; those systems have been less chemically evolved and more gas-rich than their analogues in the local universe. Hence, while bursts of star formation which accompanied mergers tend to raise metallicities of both stellar and gaseous component of the descendant's baryonic mass, the latter is partially offset by the dilution of ISM with the low-metallicity infalling gas. This might be the case with any late gaseous infall, in particular that which leaves observable footprints in ISM abundances in the Milky Way (e.g., Prodanovi{\' c} and Fields 2008).

In the course of further work, we shall attempt to find representative Local Group analogues and subsequently investigate how large fraction of their stellar mass retains habitable conditions over a prolonged period of time. This will include more realistic modelling of both irradiations by stellar explosions and galactic nuclei, as well as taking into account interstellar extinction and other local astrophysical processes. The results presented here indicate that the significant boost given to stellar metallicity by multiple mergers might at some point be counterbalanced by the increase in the irradiated mass and volume fraction as a consequence of the temporary increased SFR and supernova rate. The emerging picture is likely to resolve the tension between the results of Dayal et al. (2015) and Gobat and Hong (2016). Further work in this direction will also need to take into account weighting probability for the habitable terrestrial planet formation in such a manner to include its decline for very high metallicity regions as well (preferential formation of ``ocean worlds'', etc.).

\section{6. PRELIMINARY CONCLUSIONS}

Galactic habitability is a new and fruitful area of astrobiological research, especially within the context of theoretical and numerical astrobiology. It benefits from tremendous expansion of our knowledge of galaxies in the last half century, and it joins rather smoothly with the conventional and uncontroversial astrobiological topic of the habitability of various extrasolar planetary systems and their planets. Studying the habitability of galaxies, as the most important form of the large-scale organization of matter in our universe, makes most use of the important, and often only tacitly assumed, methodological principles of the uniformity of physical laws and universality of locally established causal relationships. On the other extreme of the complexity spectrum, both the classic rare-Earth hypothesis and its alternatives have real and tangible consequences for our SETI efforts on both Galactic and extragalactic levels (e.g., Chyba and Hand 2005; {\' C}irkovi{\' c} and Bradbury 2006; {\' C}irkovi{\' c} and Vukoti{\' c} 2008; Griffith et al. 2015). 

As we have seen in Section 2, the major factors influencing galactic habitability are still poorly understood at present. Due to the highly complex, nonlinear nature of the astrophysical processes involved in the emergence of habitability, further progress necessitates both bottom-up and top-down approaches: we need much better insight into both \textbf{(i) individual local factors} influencing habitability (such as metallicity, cosmic ray production, or nuclear activity) of external galaxies), and the ecological and biological consequences of these factors, and \textbf{(ii) complex global effects} of the interactions and processes on the level of galaxy environments (such as mergers, tidal strippings, threshings, etc., occurring in galaxy groups or clusters). It seems that even at this early stage, the field of galaxy habitability must take into account complex effects of heterogeneous galaxy environments -- something which can be done, at least in principle, within the semi-analytic merger tree approach.

Of course, the central underlying problem, which takes many forms in various aspects of exploration, stays open: to what extent is habitability of galaxies a regular, lawful consequence of physical and chemical evolution, and to what extent is it  a contingent happenstance? We need to understand and accept that there is simply no ``shortcut'' solution to this fundamental problem, at least until we gather much larger empirical database (which, when external galaxies are concerned, is likely to take a \textsl{very} long time). The best we could do is to use the Copernican assumption as the \textsl{working hypothesis}, and try to employ our best understanding of universal phenomena like galactic formation and evolution in order to obtain a set of plausible and internally consistent astrobiological {\it histories}. It is our contention, elaborated above, that the merger tree approach is by far the best way of making both sufficiently general and sufficiently precise insights into this vast and not as yet even properly glimpsed domain. 

\acknowledgements{Very helpful comments of an anonymous referee have helped improved a previous version of this manuscript. We thank Miroslav Mi{\' c}i{\' c} and Nemanja Martinovi{\' c} for useful discussions that crucially improved our understanding of the dwarf galaxies' evolution and data extraction of simulated halos. BV and MMC acknowledge financial support from the Ministry of Education, Science and Technological Development of the Republic of Serbia through the project \#ON176021 ``Visible and invisible matter in nearby galaxies: theory and observations''.}

\references

Annis, J.: 1999, \journal{J. Brit. Interpl. Soc.}, \vol{52}, 33-36.

Babcock, E. B. and Collins, J. L.: 1929, \journal{Proceedings of the National Academy of Sciences of the United States of America}, \vol{15},  623-628.

Balbi, A. and Tombesi, F.: 2017, \journal{Nature Scientific Reports}, \vol{7}, 16626 (6pp).

Balogh, M., et al.: 2004, \journal{Mon. Not. R. Astron. Soc.}, \vol{348}, 1355-1372.

Barnes, R., Meadows, V. S. and Evans, N.: 2015, \journal{Astrophys. J.}, \vol{814},  91 (11pp).

Beech, M.: 2011, \journal{Astrophys. Space Sci.}, \vol{336}(2), pp.287-302.

Begelman, M. C. and Rees, M. J.: 1976, \journal{Astrophys. J.}, \vol{814},  91 (11pp).

Bekki, K., Couch, W. J., Drinkwater, M. J. and Gregg, M. D.: 2001, \journal{The Astrophysical Journal Letters}, \vol{557},  L39-L42.

Benson, A. J. : 2010, \journal{Physics Reports}, \vol{495},  33-86.

Bora, K., Saha, S., Agrawal, S., Safonova, M., Routh, S., and Narasimhamurthy, A.: 2016, \journal{Astronomy and Computing}, \vol{17},  129-143.

Buchhave, L.A., Latham, D.W., Johansen, A., Bizzarro, M., Torres, G., Rowe, J. F., Batalha, N. M., Borucki, W.J., Brugamyer, E., Caldwell, C. and Bryson, S.T.,: 2012, \journal{Nature}, \vol{486},  375.

Cheng, J.Y., Rockosi, C.M., Morrison, H.L., Sch{\" o}nrich, R.A., Lee, Y.S., Beers, T.C., Bizyaev, D., Pan, K. and Schneider, D.P.,: 2012, \journal{Astrophys. J.}, \vol{746},  149 (23pp).

Chopra, A. and Lineweaver, C. H.: 2016, \journal{Astrobiology}, \vol{16},  7-22.

Chyba, C. F. and Hand, K.: 2005, \journal{Annu. Rev. Astron. Astrophys.}, \vol{43},  31-74.

{\' C}irkovi{\' c}, M. M. and Balbi, A.: 2019, \journal{International Journal of Astrobiology}, in press.

{\' C}irkovi{\' c}, M. M.: 2012, The Astrobiological Landscape: Philosophical Foundations of the Study of Cosmic Life, Cambridge University Press, Cambridge.

{\' C}irkovi{\' c}, M. M. and Bradbury, R. J.: 2006, \journal{New Ast.}, \vol{11},  628-639.

{\' C}irkovi{\' c} M. M. and Vukoti{\' c}, B.: 2008, \journal{Origins of Life and Evolution of Biospheres}, \vol{38},  535-547.

Clarke, J. N.: 1981, \journal{Icarus}, \vol{46},  94-96.

Clube, S. V. M.: 1978, \journal{Vistas in Astronomy}, \vol{22},  77-118.

Cockell, C.S., Bush, T., Bryce, C., Direito, S., Fox-Powell, M., Harrison, J.P., Lammer, H., Landenmark, H., Martin-Torres, J., Nicholson, N. and Noack, L.: 2016, \journal{Astrobiology}, \vol{16},  89-117.

Conway Morris, S. : 2003, Life?s Solution: Inevitable Humans in a Lonely Universe, Cambridge University Press, Cambridge.

Crick, F. H. and Orgel, L. E.: 1973, \journal{Icarus}, \vol{19},  341-346.

Crutzen, P.J. and Br{\" u}hl, C.,: 1996, \journal{Proceedings of the National Academy of Sciences}, \vol{93},  1582-1584.

Dar, A., Laor, A., and Shaviv, N. J.: 1998, \journal{Phys. Rev. Lett.}, \vol{80},  5813-5816.

Dartnell, L. R.: 2011, \journal{Astrobiology}, \vol{11},  551-582.

Dayal, P., Cockell, C., Rice, K., and Mazumdar, A.: 2015, \journal{The Astrophysical Journal Letters}, \vol{810},  L2 (5pp).

de Juan Ovelar, M., Kruijssen, J. M. D., Bressert, E., Testi, L., Bastian, N. and Cánovas, H.: 2012, \journal{Astron. Astrophys.}, \vol{546},  L1 (6pp).

Dermer, C. D. and Holmes, J. M.: 2005, \journal{Astrophys. J.}, \vol{628},  L21-L24.

Des Marais, D. J. and Walter, M. R.: 1999, \journal{Annu. Rev. Ecol. Syst.}, \vol{30},  397-420.

Des Marais, D. J., et al.: 2008, \journal{Astrobiology}, \vol{8},  715-730.

D'Souza, R. and Bell, E. F.: 2018, \journal{Nature Astronomy}, \vol{2},  737-743.

Di Stefano, R. and Ray, A.: 2016, \journal{Astrophys. J.}, \vol{827},  54 (12pp).

Domagal-Goldman, S. D., Wright, K. E., Adamala, K., Arina de la Rubia, L., Bond, J., Dartnell, L. R., Goldman, A. D., Lynch, K., Naud, M. E., Paulino-Lima, I. G. and Singer, K.: 2016, \journal{Astrobiology}, \vol{16},  561-653.

Duc, P. A., Braine, J. and Brinks, E. (eds.): 2004, \journal{Proceedings of the International Astronomical Union},  \vol{217}.

Egan, G. :1997, Diaspora, Orion/Millennium, London.

Ellis, J. and Schramm, D.N.: 1995, \journal{Proceedings of the National Academy of Sciences}, \vol{92},  235-238.

Erlykin, A. D. and Wolfendale, A. W.: 2010, \journal{Surveys in Geophysics}, \vol{31},  383-398.

Fakhouri, O. and Ma, C. P.: 2008, \journal{Mon. Not. R. Astron. Soc.}, \vol{386},  577-592.

Ferrari, F. and Szuszkiewicz, E.: 2009, \journal{Astrobiology}, \vol{9},  413-436.

Filipovic, M. D., Horner, J., Crawford, E. J., Tothill, N. F. H., and White, G. L.: 2013, \journal{Serb. Astron. J.}, \vol{187},  43-52.

Fischer, D. A. and Valenti, J.: 2005, \journal{Astrophys. J.}, \vol{622},  1102.

Forgan, D., Dayal, P., Cockell, C., and Libeskind, N.: 2017, \journal{International Journal of Astrobiology}, \vol{16},  60-73.

Franck, S., Block, A., von Bloh, W., Bounama, C., Garrido, I. and Schellnhuber, H. -J.: 2001, \journal{Naturwiss.}, \vol{88},  416?426.

Gehrels, N., Laird, C.M., Jackman, C.H., Cannizzo, J.K., Mattson, B.J. and Chen, W.: 2003, \journal{Astrophys. J.}, \vol{585},  1169-1176.

Gobat, R. and Hong, S. E.: 2016, \journal{Astron. Astrophys.}, \vol{592},  A96 (10pp).

Gonzalez, G. : 2005, \journal{Origin of Life and Evolution of the Biosphere}, \vol{35},  555-606.

Gonzalez, G., Brownlee, D., and Ward, P.: 2001, \journal{Icarus}, \vol{152},  185-200.

Gould, S. J.: 1989, Wonderful Life: The Burgess Shale and the Nature of History, W. W. Norton, New York.

Gowanlock, M. G., Patton, D. R., and McConnell, S. M.: 2011, \journal{Astrobiology}, \vol{11},  855-873.

Griffith, R. L., Wright, J. T., Maldonado, J., Povich, M. S., Sigurdsson, S., and Mullan, B.: 2015, \journal{The Astrophysical Journal Supplement Series}, \vol{217},  article id. 25 (34pp).

Hammer, F., Puech, M., Chemin, L., Flores, H., and Lehnert, M. D.: 2007, \journal{Astrophys. J.}, \vol{662},  322-334.

Hanslmeier, A. : 2009, Habitability and Cosmic Catastrophes, Springer, Berlin.

Haqq-Misra, J., Kopparapu, R., and Wolf, E.: 2018, \journal{International Journal of Astrobiology}, \vol{17},  77-86.

Heitsch, F. and Putman, M. E.: 2009, \journal{Astrophys. J.}, \vol{698},  1485-1496.

Heller, R.: 2012, \journal{Astron. Astrophys.}, \vol{545},  L8 (4pp).

Heller, R., Williams, D., Kipping, D., Limbach, M. A., Turner, E., Greenberg, R., Sasaki, T., Bolmont, E., Grasset, O., Lewis, K., and Barnes, R.: 2014, \journal{Astrobiology}, \vol{14},  798-835.

Hoffman, P. H., Kaufman, A. J., Halverson, G. P., and Schrag, D. P.: 1998, \journal{Astrobiology}, \vol{14},  798-835.

Hogan, C. J.: 2000, \journal{Reviews of Modern Physics}, \vol{72},  1149-61.

Hopkins, A. M. and Beacom, J. F.: 2006, \journal{Astrophys. J.}, \vol{651},  142.

Horneck, G. and Rettberg, P. (eds.): 2007, Complete Course in Astrobiology, Wiley-VCH, Weinheim.

Hoyle, F. and Hoyle, G.: 1973, The Inferno, William Heinemann, London.

Humphries, C. J. and Parenti, L. R.: 1999, Cladistic biogeography, OUP, Oxford.

Jagadeesh, M.K., Gudennavar, S.B., Doshi, U. and Safonova, M.: 2017, \journal{Astrophys. Space Sci.}, \vol{362}(8),  146 (13pp).

Jagadeesh, M.K., Roszkowska, M. and Kaczmarek, \L.: 2018, \journal{Life sciences in space research}, \vol{19},  13-16. 

Jiang, F. and van den Bosch, F. C.: 2014, \journal{Mon. Not. R. Astron. Soc.}, \vol{440},  193-207.

Jovanovi{\' c}, M.: 2017, \journal{Mon. Not. R. Astron. Soc.}, \vol{469},  3564-3575.

Kirby, E. N., Cohen J. G., Guhathakurta, P., Cheng, L., Bullock, J. S., and Gallazzi, A.: 2013, \journal{Astrophys. J.}, \vol{779},  102.

Kokaia, G. and Davies, M. B.: 2019, \journal{Mon. Not. R. Astron. Soc.},  in press (preprint arXiv:1903.08026v2).

Kragh, H.: 1996, Cosmology and Controversy, Princeton University Press, Princeton.

Krasovsky, V. I. and Shklovsky, I. S.: 1957, \journal{Dokl. Akad. Nauk SSSR}, \vol{116}, 197.

Lacey, C. and Cole, S.: 1993, \journal{Mon. Not. R. Astron. Soc.}, \vol{262},  627-649.

Laster, H., Tucker, W. H., and Terry, K. D.: 1968, \journal{Science}, \vol{160},  1138-1139.

Laughlin, G. and Adams, F. C.: 2000, \journal{Icarus}, \vol{145},  614-627.

Li, Y. and Zhang, B.: 2015, \journal{Astrophys. J.}, \vol{810}, 41 (7pp).

Lineweaver, C. H.: 2001, \journal{Icarus}, \vol{151},  307-313.

Lineweaver, C. H., Fenner, Y., and Gibson, B. K.: 2004, \journal{Science}, \vol{303},  59-62.

Lingam, M. and Loeb, A.: 2018, \journal{Astrobiology}, \vol{19},  28-39.

Lunine, J. I.: 2009, \journal{Proceedings of the American Philosophical Society}, \vol{153},  403-418.

Marinho, F., Paulucci, L., and Galante, D.: 2014, \journal{International Journal of Astrobiology}, \vol{13},  319-323.

Martinovi{\' c}, N., and Micic, M.: 2017, \journal{Mon. Not. R. Astron. Soc.}, \vol{470},  4015-4025.

Mateo, M. L.: 1998, \journal{Annual Review of Astronomy and Astrophysics}, \vol{36},  435?506.

McConnachie, A. W.: 2012, \journal{The Astronomical Journal}, \vol{144},  4 (36pp).

McKay, C.P., Anbar, A.D., Porco, C. and Tsou, P.,: 2014, \journal{Astrobiology}, \vol{14},  352-355.

Melott, A. L. and Thomas, B. C.: 2011, \journal{Astrobiology}, \vol{11},  343-361.

Micic, M., Martinovi{\' c}, N., and Sinha, M.: 2016, \journal{Mon. Not. R. Astron. Soc.}, \vol{461},  3322-3335.

Mitri, G., Postberg, F., Soderblom, J.M., Wurz, P., Tortora, P., Abel, B., Barnes, J.W., Berga, M., Carrasco, N., Coustenis, A. and de Vera, J.P.P., :2018, \journal{Planetary and Space Science}, \vol{155},  73-90.

Moeller, R., Reitz, G., Berger, T., Okayasu, R., Nicholson, W. L. and Horneck, G.: 2010, \journal{Astrobiology}, \vol{10},  509-521.

Nelson, D., Pillepich, A., Genel, S., Vogelsberger, M., Springel, V., Torrey, P., Rodriguez-Gomez, V., Sijacki, D., Snyder, G.F., Griffen, B., and Marinacci, F.: 2015, \journal{Astronomy and Computing}, \vol{13},  12-37.

Pavlov, A. A., Toon, O. B., Pavlov, A. K., Bally, J., and Pollard, D.: 2005, \journal{Geophys. Res. Lett.}, \vol{32},  L03705[4].

Pedicelli, S.E.E., Bono, G., Lemasle, B., François, P., Groenewegen, M., Lub, J., Pel, J.W., Laney, D., Piersimoni, A., Romaniello, M. and Buonanno, R.: 2009, \journal{Astron. Astrophys.}, \vol{504},  81-86.

Peeples, M. S., Pogge, R. W. and Stanek, K. Z.: 2008, \journal{Astrophys. J.}, \vol{685},  904-914.

Poulton, R. J., Robotham, A. S., Power, C., and Elahi, P. J.: 2018, \journal{Publications of the Astronomical Society of Australia}, \vol{35},  e042 (14pp).

Press, W. H. and Schechter, P.: 1974, \journal{Astrophys. J.}, \vol{187},  425-438.

Prodanovi{\' c}, T. and Fields, B. D.: 2008, \journal{Journal of Cosmology and Astroparticle Physics}, \vol{9}, 3 (22pp).

Rodriguez-Gomez, V., Genel, S., Vogelsberger, M., Sijacki, D., Pillepich, A., Sales, L.V., Torrey, P., Snyder, G., Nelson, D., Springel, V. and Ma, C. P.: 2015, \journal{Mon. Not. R. Astron. Soc.}, \vol{449},  49-64.

Rodr{\' i}guez-Mozos, J. M. and Moya, A.: 2017, \journal{Mon. Not. R. Astron. Soc.}, \vol{471},  4628-4636.

Saha, S., Basak, S., Safonova, M., Bora, K., Agrawal, S., Sarkar, P. and Murthy, J.: 2018, \journal{Astronomy and computing}, \vol{23},  141-150.

Sanders, R.H. and Prendergast, K.H.: 1974, \journal{Astrophys. J.}, \vol{188},  pp.489-500.

Scalo, J. and Wheeler, J. C.: 2002, \journal{Astrophys. J.}, \vol{566},  723-737.

Schlaufman, K.C. and Laughlin, G.: 2011, \journal{Astrophys. J.}, \vol{738},  177.

Seager, S.: 2013, \journal{Science}, \vol{340},  577-581.

Shields, A. L., Ballard, S. and Johnson, J. A.: 2016, \journal{Physics Reports}, \vol{663},  1-38.

Shock, E. L. and Holland, M. E.: 2007, \journal{Astrobiology}, \vol{7},  839-851.

Simpson, G. G.: 1968, \journal{Science}, \vol{162},  140-141.

Stanway, E. R., Hoskin, M. J., Lane, M. A., Brown, G. C., Childs, H. J. T., Greis, S. M. L., and Levan, A. J.: 2018, \journal{Mon. Not. R. Astron. Soc.}, \vol{475},  1829-1842.

Stojkovi{\' c}, N., Vukoti{\' c}, B., Martinovi{\' c}, N., {\' C}irkovi{\' c}, M. M., and Micic, M. : 2019, \journal{Mon. Not. R. Astron. Soc.}, submitted.

Su, M., Slatyer, T. R., and Finkbeiner, D. P.: 2010, \journal{Astrophys. J.}, \vol{724},  1044?1082.

Sullivan, M., Le Borgne, D., Pritchet, C.J., Hodsman, A., Neill, J.D., Howell, D.A., Carlberg, R.G., Astier, P., Aubourg, E., Balam, D. and Basa, S.: 2006, \journal{Astrophys. J.}, \vol{648},  868-883.

Suthar, F. and McKay, C. P.: 2012, \journal{International Journal of Astrobiology}, \vol{11},  157-161.

Tadross, A. L.: 2003, \journal{New Astronomy}, \vol{8},  737-744.

Thomas, B. C.: 2009, \journal{International Journal of Astrobiology}, \vol{8}, 183.

Thomas, B. C., Jackman, C. H., Melott, A. L., Laird, C. M., Stolarski, R. S., Gehrels, N., Cannizzo, J. K., Hogan, D. P.: 2005, \journal{Astrophys. J.}, \vol{622}, L153.

Thomas, B. C., Melott, A. L., Fields, B. D. and Anthony-Twarog, B. J. : 2008, \journal{Astrobiology}, \vol{8},  9-16.

Thomas, B. C. and Melott, A. L.: 2006, \journal{New Journal of Physics}, \vol{8}, 120.

Turner, M.S.: 2018, \journal{Foundations of Physics}, \vol{48},  1261-1278.

van Bueren, H. G.: 1978, \journal{Astron. Astrophys.}, \vol{70},  707-717.

Vermeij, G. J.: 2006, \journal{PNAS}, \vol{103},  1804-1809.

Vogelsberger, M., Sijacki, D., Kere{\v s}, D., Springel, V., and Hernquist, L.: 2012, \journal{Mon. Not. R. Astron. Soc.}, \vol{425},  3024?3057.

Vogelsberger, M., Genel, S., Springel, V., Torrey, P., Sijacki, D., Xu, D., Snyder, G., Nelson, D., and Hernquist, L.: 2014, \journal{Mon. Not. R. Astron. Soc.}, \vol{425},  3024?3057.

Vukoti{\' c}, B.,: 2017, in ``Habitability of the Universe Before Earth'', eds.  R. Gordon and A. A. Sharov, Elsevier B.V., Amsterdam, 174-197.

Vukoti{\' c}, B. and {\' C}irkovi{\' c}, M. M.: 2008, \journal{Serb. Astron. J.}, \vol{176},  71-79.

Vukoti{\' c}, B., Steinhauser, D., Martinez-Aviles, G., {\' C}irkovi{\' c}, M. M., Micic, M., and Schindler, S.: 2016, \journal{Mon. Not. R. Astron. Soc.}, \vol{459},  3512-3524.

Wallace, A. R.: 1876, The Geographical Distribution of Animals; With A Study of the Relations of Living and Extinct Faunas as Elucidating the Past Changes of the Earth's Surface, Macmillan \& Co., London.

Wallace, A. R.: 1903, Man's Place in the Universe; A Study of the Results of Scientific Research in Relation to the Unity or Plurality of Worlds, Chapman \& Hall, London.

Ward, P. D. and Brownlee, D.: 2000, Rare Earth: Why Complex Life Is Uncommon in the Universe, Springer, New York.

Weinberg, S.: 2008, Cosmology, Oxford University Press, Oxford.

Wright, J. T., Mullan, B., Sigurdsson, S., and Povich, M. S.: 2014, \journal{Astrophys. J.}, \vol{792},  id. 26 (16pp).

Wright, J. T., Griffith, R. L., Sigurdsson, S., Povich, M. S., and Mullan, B.: 2014, \journal{Astrophys. J.}, \vol{792},  id. 27 (12pp).

Yin, J., Hou, J. L., Prantzos, N., Boissier, S., Chang, R. X., Shen, S.Y., and Zhang, B.: 2009, \journal{Astron. Astrophys.}, \vol{505},  497-508.

Zackrisson, E.,  Calissendorff, P., Asadi, S., and Nyholm, A.: 2015, \journal{Astrophys. J.}, \vol{810},  article id. 23 (12 pp).

Zhang, J., Fakhouri, O. and Ma, C. P.: 2008, \journal{Mon. Not. R. Astron. Soc.}, \vol{389},  1521-1538.

\endreferences

}
%\end{multicols}

%\vfill\eject

{\ }

\end{document}